\documentclass[lettersize,journal]{IEEEtran}
\IEEEoverridecommandlockouts

\usepackage{amssymb}
\usepackage[cmex10]{amsmath}
\usepackage{stfloats}
\usepackage{graphicx}
\usepackage{subfigure}
\usepackage{tabularx}
\usepackage{epsfig,epsf,color,balance,cite}
\usepackage{verbatim}
\usepackage{url}
\usepackage{bm}
\usepackage{booktabs}

\usepackage{amsmath}
\allowdisplaybreaks[2]
\newtheorem{theorem}{Theorem}
\newtheorem{lemma}{Lemma}

\usepackage{algorithm}
\usepackage{algorithmic}

\usepackage{color}
\definecolor{myc1}{rgb}{0,0,0}

\begin{document}

\title{Energy-Efficient Probabilistic Semantic Communication over Space-Air-Ground\\ Integrated Networks
}

\author{Zhouxiang Zhao, 
        Zhaohui Yang, 
        Mingzhe Chen, 
        Zhaoyang Zhang,~\IEEEmembership{Senior Member,~IEEE,}\\
        Wei Xu,~\IEEEmembership{Senior Member,~IEEE,}
        and Kaibin Huang,~\IEEEmembership{Fellow,~IEEE}
\thanks{Z. Zhao, Z. Yang, and Z. Zhang are with College of Information Science and Electronic Engineering, Zhejiang University, and also with Zhejiang Provincial Key Laboratory of Info. Proc., Commun. \& Netw. (IPCAN), Hangzhou, 310027, China (e-mails: \{zhouxiangzhao, yang\_zhaohui, ning\_ming\}@zju.edu.cn).}
\thanks{M. Chen is with Department of Electrical and Computer Engineering and Institute for Data Science and Computing, University of Miami, Coral Gables, FL, 33146, USA (e-mail: mingzhe.chen@miami.edu).}
\thanks{W. Xu is with National Mobile Communications Research Laboratory, Southeast University, Nanjing, 211189, China (e-mail: wxu@seu.edu.cn).}
\thanks{K. Huang is with the Department of Electrical and Electronic Engineering, The University of Hong Kong, Hong Kong, SAR, China (e-mail: huangkb@eee.hku.hk).}
}



\maketitle

\begin{abstract}
Space-air-ground integrated networks (SAGINs) are emerging as a pivotal element in the evolution of future wireless networks. Despite their potential, the joint design of communication and computation within SAGINs remains a formidable challenge. In this paper, the problem of energy efficiency in SAGIN-enabled probabilistic semantic communication (PSC) system is investigated. In the considered model, a satellite needs to transmit data to multiple ground terminals (GTs) via an unmanned aerial vehicle (UAV) acting as a relay. During transmission, the satellite and the UAV can use PSC technique to compress the transmitting data, while the GTs can automatically recover the missing information. The PSC is underpinned by shared probability graphs that serve as a common knowledge base among the transceivers, allowing for resource-saving communication at the expense of increased computation resource. Through analysis, the computation overhead function in PSC is a piecewise function with respect to the semantic compression ratio. Therefore, it is important to make a balance between communication and computation to achieve optimal energy efficiency. The joint communication and computation problem is formulated as an optimization problem aiming to minimize the total communication and computation energy consumption of the network under latency, power, computation capacity, bandwidth, semantic compression ratio, and UAV location constraints. To solve this non-convex non-smooth problem, we propose an iterative algorithm where the closed-form solutions for computation capacity allocation and UAV altitude are obtained at each iteration. Numerical results show the effectiveness of the proposed algorithm.
\end{abstract}

\IEEEpeerreviewmaketitle

\section{Introduction}
\IEEEPARstart{I}{n} the era of rapid advancements in satellite and unmanned aerial vehicle (UAV) technology, the potential of space-air-ground integrated networks (SAGINs) to revolutionize future communication systems is becoming increasingly evident \cite{10080878}. SAGIN covers satellites, aerial platforms, and terrestrial nodes, making it a multi-level three-dimensional (3D) network \cite{8368236}. This 3D characteristic gives SAGINs global broadband coverage, which is a major demand in future wireless networks \cite{9177315}. Moreover, with the exponential growth in computation capability of ground, aerial and space platforms, SAGINs are able to provide computing-intensive services, such as semantic communication \cite{10445211}. These advantages make SAGIN a promising key technology for future mobile networks.

Satellite is a key component in SAGINs, and satellite communication has several unique advantages compared to conventional terrestrial communication \cite{9210567}. Firstly, it provides global coverage, enabling communication in remote and isolated areas where traditional infrastructure is impractical \cite{9261433}. Satellite networks also offer high reliability and resilience, as they are less susceptible to natural disasters or localized disruptions that can affect terrestrial communication systems \cite{8434289}. Additionally, satellite communication supports high data transfer rates, facilitating the transmission of large volumes of information for applications like broadband internet, television broadcasting, and secure military communications \cite{9955992,9848831}.
The advent of Starlink, a satellite internet constellation developed by SpaceX, exemplifies the remarkable advancements in satellite communication \cite{9393372}. Starlink leverages a massive network of low Earth orbit (LEO) satellites to provide high-speed, low-latency internet access globally. The deployment of Starlink showcases the potential for satellite communication to revolutionize global internet infrastructure, offering a scalable and flexible solution to meet the growing demands for high-speed connectivity in a wide range of applications.

UAVs are essential in improving the convenience of SAGINs by offering a unique combination of multi-functionality, high mobility, ease of deployment, and cost effectiveness \cite{8533634}. As a result, UAV-assisted communication stands out as an indispensable technology in the evolution of SAGINs.
The growing interest in UAV-enabled wireless communication networks has prompted researchers to investigate various aspects of this field. The authors of \cite{8586877} went into the intricacies of beamforming and power allocation. Their investigation, based on the assumption of a UAV following a circular trajectory and employing a decode-and-forward relaying strategy, aimed at maximizing the instantaneous data rate. In \cite{8941314}, the researchers explored a UAV-enabled wireless communication system with energy harvesting. This investigation considered the role of the UAV in transferring energy to users, operating in either half-duplex or full-duplex mode, while users harnessed energy for subsequent data transmission to the UAV. Furthermore, Nguyen et al. proposed a real-time resource allocation algorithm in \cite{8457275}. Their focus was on maximizing energy efficiency through joint optimization of energy harvesting time and power control in a device-to-device communication framework integrated with UAV technology. Moreover, the work presented in \cite{8764580} addressed the problem of sum power minimization within a mobile edge computing network incorporating multiple UAVs. This effort involved the simultaneous optimization of user association, power control, computation capacity allocation, and location planning.

The fast development in satellite communication and UAV-assisted communication has greatly stimulated the potential of SAGINs. However, the aforementioned works \cite{9261433,8434289,9955992,9848831,9393372,8533634,8586877,8941314,8457275,8764580} all used conventional communication schemes. By employing the emerging semantic communication, we could unlock the potential of SAGINs even further.

Semantic communication represents a novel paradigm that transcends traditional bit transmission methods \cite{9955525,10550151,9679803,9955312,9530497}. Unlike conventional communication, which focuses on transmitting identical bit streams without considering the underlying message meaning, semantic communication prioritizes conveying the actual meaning of the message \cite{zhou2024feature}. This semantic content is typically much smaller than the original bit data \cite{10024766}. Taking advantage of the rapid advancements in artificial intelligence, we can use this technology to accurately extract semantic information from the original data \cite{10233741,9398576}. By targeting the extraction of essential meaning, semantic communication has the capability to eliminate redundant data, particularly those less relevant to the receiver's task \cite{zhao2024prob}. Consequently, the communication process becomes more energy efficient, exhibits higher spectral efficiency, and ensures greater reliability \cite{10183794}.

Semantic communication has emerged as a vibrant area of research within the communication domain, witnessing several impactful studies. In \cite{9832831}, researchers introduced a semantic communication framework designed for the transmission of textual data. The framework utilized a knowledge graph composed of semantic triples to model the semantic information. Moreover, \cite{9953316} explored semantic-oriented speech transmission. They devised an end-to-end deep learning-based transceiver capable of extracting and encoding semantic information from input speech spectrums at the transmitter. At the receiver end, the system decoded this information to produce corresponding transcriptions, comprising a compact set of semantically irrelevant details essential for speech reconstruction. Furthermore, \cite{9953076} proposed a reinforcement learning-based adaptive semantic coding strategy for image data, transcending pixel-level encoding. Introducing the concept of semantic representation units that encompass category, spatial arrangement, and visual features, the authors designed a convolutional semantic encoder. This encoder effectively extracted semantic concepts, paving the way for adaptive and efficient image encoding beyond traditional pixel-level methodologies.

The aforementioned observations highlight the significant computation demands associated with semantic communication \cite{e26050394}. Specifically, the extraction of semantic meaning from messages requires substantial computation resources, resulting in a reduction of the transmitted data volume and a subsequent decrease in the demand for communication resources \cite{9763856,10032275}. Notably, within SAGINs, communication resources are more valuable when compared to traditional terrestrial communication frameworks \cite{10440193}. Consequently, the integration of semantic communication in SAGINs is considered prudent, as it helps economize on communication resources, despite the associated computation overhead. The increasing computation capabilities of space, air, and ground platforms within SAGINs contribute to a growing abundance of computation resources, thereby strengthening the foundation for effective semantic communication. This connection underscores an interdependence relationship between SAGIN and semantic communication, emphasizing their mutual reinforcement and substantiating their collective capacity to optimize information exchange within resource-constrained environments.

This paper explores the integration of SAGIN and semantic communication. We propose a SAGIN-enabled probabilistic semantic communication (PSC) framework that combines the benefits of SAGIN and semantic communication. To the best of our knowledge, this is the first work that examines semantic communication resource allocation problem in SAGINs. The key contributions are summarized as follows:
\begin{itemize}
\item We propose a SAGIN-enabled PSC framework in which a satellite in the space transmits data to multiple ground terminals (GTs) and uses a UAV in the air as a relay. During the transmission, the satellite and the UAV can choose to perform semantic compression to reduce the data size of the message using PSC technique, while the GTs can reconstruct the compressed information using the shared probability graphs. Although this semantic compression can save communication resource, it inevitably incurs additional computation costs. The computation overhead function in PSC system is modeled as a piecewise function with respect to the semantic compression ratio. Thus, a strategic balance between communication and computation in the system is required to achieve optimal performance.
\item We formulate a joint communication and computation optimization problem whose goal is minimizing the total communication and computation energy consumption of the system while considering latency, power, computation capacity, bandwidth, semantic compression ratio, and UAV location constraints.
\item To solve this non-convex non-smooth problem, we propose an iterative algorithm where the closed-form solutions for computation capacity allocation and UAV altitude are obtained at each iteration. Numerical results show the effectiveness of the proposed algorithm.
\end{itemize}

The remainder of this paper is organized as follows. The system model and problem formulation are described in Section \ref{smpf}. The algorithm design is presented in Section \ref{ad}. Simulation results are analyzed in Section \ref{sra}. Conclusions are drawn in Section \ref{c}.

\section{System Model and Problem Formulation}\label{smpf}

\begin{figure}[t]
\centering
\includegraphics[width=\linewidth]{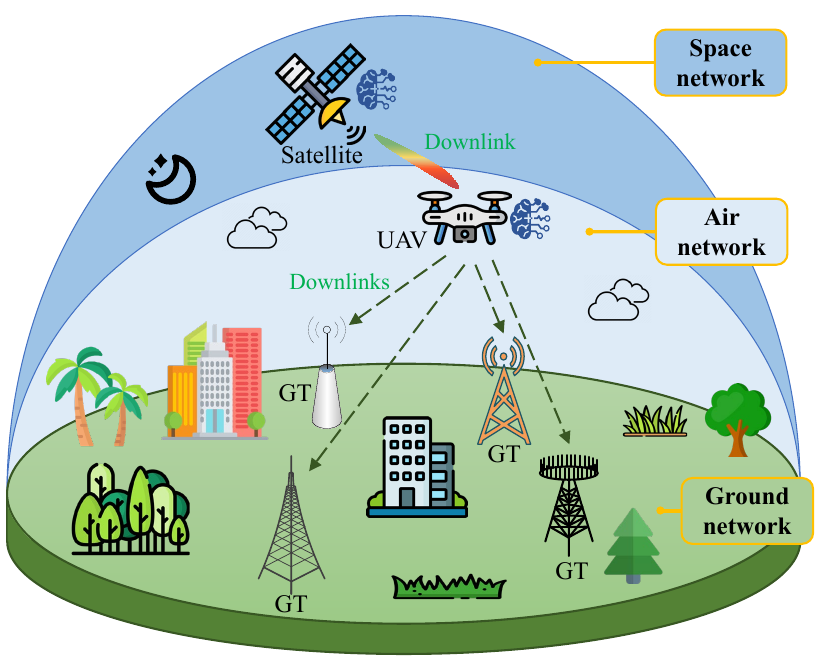}
\caption{Illustration of the considered SAGIN-enabled PSC system.}
\label{fg.s}
\end{figure}

Consider a downlink transmission scenario that a satellite in the space needs to transmit data to multiple GTs on the ground and uses a UAV in the air as a relay, as shown in Fig.~\ref{fg.s}. The set of GTs is represented by $\mathcal{K}=\{1,2,\cdots,K\}$, and the data that need to be transmitted to GT $k$ is denoted by $\mathcal{D}_k$. Due to limited wireless resources, the satellite or UAV needs to take advantage of its computation capability to extract the small-sized semantic information $\mathcal{C}_k$ from original data $\mathcal{D}_k$ to reduce data size, thus saving communication resources. In the considered model, semantic communication is enabled by shared probability graphs between the satellite, UAV, and GTs.

\subsection{Semantic Communication Model}
In the considered PSC model, we assume that the transmitter (satellite) has substantial semantic triples \cite{10118916} to transmit. A typical semantic triple can be represented by 
\begin{equation}
    \varepsilon = (h, r, t),
\end{equation}
where $h$ is the head entity, $t$ denotes the tail entity, and $r$ represents the relation between $h$ and $t$. For instance, (\textit{Tree, on, Grass}) is a semantic triple, where ``\textit{Tree}'' is the head entity, ``\textit{Grass}'' is the tail entity, and ``\textit{on}'' is the relation, and this semantic triple basically means ``the tree grows on the grass''. These semantic triples can be generated from extensive textual/audio/picture/video data using deep neural networks \cite{9039685,9446853,10226302,10214643,9961954,8788525}. As demonstrated by the instance, the semantic triple is a data representation approach with high information density. It uses a small number of bits to store a relatively large amount of information. Although the semantic triple already uses a small amount of data, we propose a probabilistic approach that can compress it even further.

\begin{figure}[t]
\centering
\includegraphics[width=0.9\linewidth]{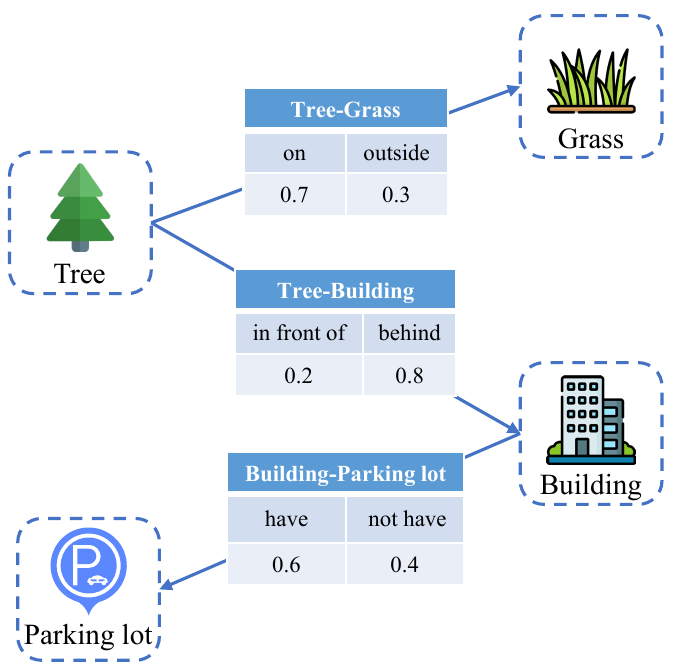}
\caption{Illustration of the probability graph in the PSC system.}
\label{fg.pg}
\end{figure}

In a traditional semantic triple, the relation is fixed. If there are substantial semantic triples, we can use them as samples to generate a probability graph that stores the statistical information of these semantic triples. These sample semantic triples used to construct the probability graph can be historical data between the transceivers. The probability graph consists of numerous semantic quadruples as shown in Fig.~\ref{fg.pg}. The semantic quadruple extends the conventional semantic triple with an additional dimension of relation probability, which reflects the probability of a particular relation occurring under the given condition of a fixed head entity and tail entity. The details of probability graph construction can be found in \cite{10333452}.

The probability graph is shared between the semantic transceivers as a knowledge base. When the transmitter needs to send a set of semantic triples to the receiver, the probability graph can be used for further semantic compression. Using the statistical data from the probability graph, a multi-dimensional conditional probability matrix can be calculated. This matrix indicates the probability of a specific semantic triple being valid, given that certain other triples are valid. This allows the removal of relation in the semantic triple before transmission, resulting in data compression. After the receiver receives the compressed information, it can recover the missing relation according to the shared probability graph. For example, assuming that the probability graph in Fig.~\ref{fg.pg} is shared by the transceivers, if the transmitter needs to send a semantic triple (\textit{Tree, behind, Building}) to the receiver, then according to the probability graph, it only needs to send (\textit{Tree, $\varnothing$, Building}) where the relation is omitted. After receiving the compressed information, the receiver can automatically recover the relation ``\textit{behind}'' based on the shared probability graph, because it is of higher probability. However, it is important to emphasize that achieving a smaller data size requires a lower semantic compression ratio, which requires higher-dimensional conditional probabilities. This decrease in the semantic compression ratio results in a higher computation overhead, which represents a trade-off between communication and computation in the PSC network.

Within the considered SAGIN-enabled PSC framework, each GT has a local probability graph that stores statistical information about its historical data. These probability graphs are shared with the satellite and the UAV. The satellite has data $\mathcal{D}_k$ for GT $k$, and these data can be remote sensing results in the form of semantic triples. With the above PSC technique, the satellite or UAV can perform semantic compression to further compress the data $\mathcal{D}_k$. After GT $k$ receives the compressed data denoted by $\mathcal{C}_k$, it can automatically recover missing information using the shared probability graph.
Denote the number of bits in original data $\mathcal{D}_k$ by $D_k$ and the number of bits in compressed data $\mathcal{C}_k$ by $C_k$, then the semantic compression ratio for GT $k$ can be defined as
\begin{equation}
    \rho_k=\frac{C_k}{D_k}.
\end{equation}
A lower semantic compression ratio indicates a more compact compression.

\subsection{Computation Model}
The PSC technique can save communication resources through semantic compression. However, this process inevitably costs additional computation resources. Intuitively, a lower semantic compression ratio corresponds to a larger computation consumption, as it requires more compression.

According to equation (20) in \cite{ZHAO2024107055}, the computation overhead for the considered PSC technique can be written as
\begin{equation}\label{eq.co}
    O_k\left(\rho_k\right)=\begin{cases}
        A_{k1}\rho_k +B_{k1}, &C_{k1}< \rho_k \leq 1,\\
        A_{k2}\rho_k +B_{k2}, &C_{k2}< \rho_k \leq C_{k1},\\
        \vdots\\
        A_{kD}\rho_k +B_{kD}, &C_{kD}\leq \rho_k \leq C_{k(D-1)},
    \end{cases}
\end{equation}
where $A_{kd}<0$ denotes the slope, $B_{kd}>0$ represents the constant term, and $C_{kd}$ stands for the boundary for each segment $d\in\mathcal{D}^\mathrm{s}=\left\{1, \cdots, D\right\}$. These parameters depend on the properties of the probability graph corresponding to different GTs.

\begin{figure}[t]
\centering
\includegraphics[width=\linewidth]{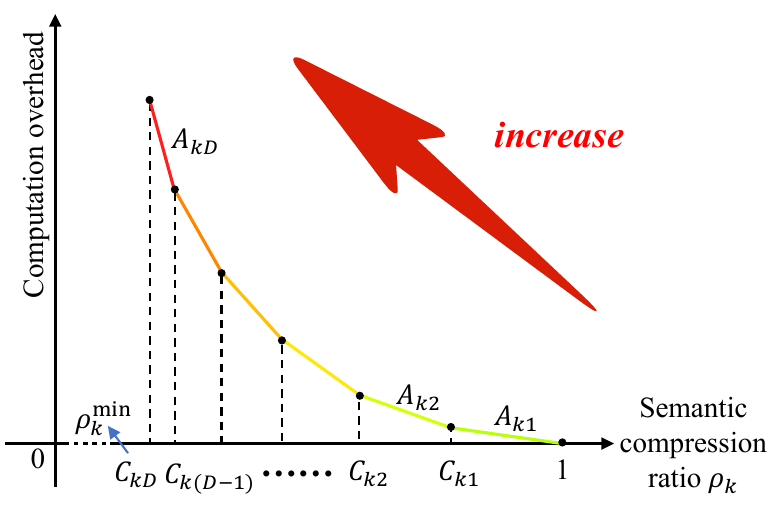}
\caption{Illustration of the relationship between semantic compression ratio and computation overhead.}
\label{fg.co}
\end{figure}

From equation~\eqref{eq.co}, we observe that the computation overhead for GT $k$, denoted by $O_k(\rho_k)$, manifests as a piecewise function with respect to its semantic compression ratio, $\rho_k$. Fig.~\ref{fg.co} visually depicts the behavior of the $O_k(\rho_k)$ function. This function exhibits a segmented structure with $D$ distinct levels. This segmentation arises from the inherent characteristic of the semantic compression process, which employs multiple hierarchical levels of conditional probabilities. Each level contributes a distinct linear expression to the overall computation overhead. Furthermore, a discernible decrease in the slope magnitude of $O_k(\rho_k)$ is evident across these discrete segments. This phenomenon can be attributed to the utilization of lower-dimensional conditional probabilities at higher semantic compression ratios, resulting in reduced computation demands. On the contrary, as $\rho_k$ diminishes, the need for higher-dimensional information arises. Consequently, the computation overhead intensifies with increasing information dimensionality. Each transition within the piecewise function $O_k(\rho_k)$ signifies the activation of a higher level of probabilistic information.

\subsection{Network Model}

\begin{figure}[t]
\centering
\includegraphics[width=\linewidth]{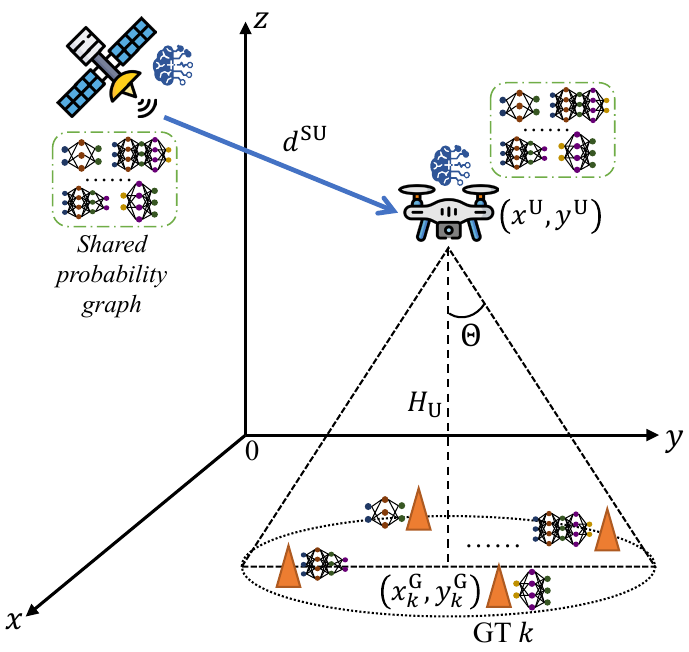}
\caption{The considered SAGIN-enabled PSC network.}
\label{fg.nm}
\end{figure}

In the considered SAGIN-enabled PSC system, there is one satellite, one UAV, and $K$ GTs. The satellite and the UAV are equipped with computation capacity and are shared with probability graphs of all GTs. The UAV can hover in the air. Fig.~\ref{fg.nm} illustrates the considered network.

In the ground network, the horizontal and vertical location of GT $k$ can be denoted by
\begin{equation}
    \mathbf{L}_k^\mathrm{G}=\left(x_k^\mathrm{G},y_k^\mathrm{G}\right),
\end{equation}
and the height of each GT is approximated to be zero compared to the height of UAV and satellite.

In the air network, the horizontal and vertical location of the UAV can be denoted by
\begin{equation}
    \mathbf{L}^\mathrm{U}=\left(x^\mathrm{U},y^\mathrm{U}\right),
\end{equation}
and the height of the UAV is represented by $H_\mathrm{U}$. Therefore, the distance between the UAV and GT $k$ can be calculated as
\begin{equation}
    d_k^\mathrm{UG}=\left(\left\|\mathbf{L}^\mathrm{U}-\mathbf{L}_k^\mathrm{G}\right\|^2+H_\mathrm{U}^2\right)^{\frac{1}{2}},
\end{equation}
where $\left\|\cdot\right\|$ is the Euclidian norm.

In the space network, we denote the distance between the satellite and the UAV by $d^\mathrm{SU}$. Since the altitude of the satellite is very high, the positional change of the UAV has little effect on $d^\mathrm{SU}$. For simplification, we disregard the impact of the UAV's positional changes on the distance between the satellite and the UAV.

The satellite transmits data to GT $k$ indirectly with the UAV as a relay. During the transmission process, the satellite or UAV can choose to perform semantic compression using the PSC technique to reduce the consumption of communication resources. We use $a_k^\mathrm{S}$ to indicate the computation state of the satellite. If the satellite does the semantic compression for GT $k$, we have $a_k^\mathrm{S}=1$; otherwise, we have $a_k^\mathrm{S}=0$. Similarly, we use $a_k^\mathrm{U}$ to indicate the computation state of the UAV. If the UAV does the semantic compression for GT $k$, we have $a_k^\mathrm{U}=1$; otherwise, we have $a_k^\mathrm{U}=0$. Since the semantic compression for each GT is required at most once, the following constraint can be obtained:
\begin{equation}
    a_k^\mathrm{S}+a_k^\mathrm{U}\leq 1,
\end{equation}
which indicates that the semantic compression for each GT is either conducted by the satellite, by the UAV, or by neither of them.

\subsection{Transmission Model}
In the considered downlink transmission scenario, the satellite sends data to the UAV, which then sends data to the GTs.

\subsubsection{Satellite to UAV}
Different from traditional terrestrial communication, the satellite communication channel is subject to various factors, including space propagation fading, atmospheric absorption fading, rain attenuation, among others. For simplification, we model the satellite-to-UAV wireless channel coefficient as
\begin{equation}
    \left|h_\mathrm{SU}\right|=\frac{\sqrt{\delta_\mathrm{S}}\lambda^\mathrm{SU}}{4\pi d^\mathrm{SU}},
\end{equation}
where $\delta_\mathrm{S}$ is the beam gain, and $\lambda^\mathrm{SU}$ denotes the wavelength of the satellite-to-UAV transmission wave.

Consequently, the downlink transmission rate between the satellite and the UAV can be given by
\begin{equation}
    r_\mathrm{SU}=B_\mathrm{SU}\log_2\left(1+\frac{\left|h_\mathrm{SU}\right|^2 P_\mathrm{S}}{B_\mathrm{SU}N_0}\right),
\end{equation}
where $B_\mathrm{SU}$ denotes the bandwidth of the satellite-to-UAV system, $P_\mathrm{S}$ represents the transmit power of the satellite, and $N_0$ is the power spectral density of additive white Gaussian noise (AWGN).

\subsubsection{UAV to GT}
We assume that the UAV is outfitted with a directional antenna featuring adjustable beamwidth, while each GT is equipped with an omnidirectional antenna possessing unit gain. The azimuth and elevation half-power beamwidths of the UAV's antenna are equivalent and are denoted by $2\Theta\in(0,\pi)$. Conforming to [\cite{balanis2016antenna}, Eqs. (2-51)], the antenna gain within the azimuth angle $\theta$ and elevation angle $\phi$ can be expressed as
\begin{equation}
    G=\begin{cases}
        \frac{G_0}{\Theta^2}, &\text{if }0\leq\theta\leq\Theta\text{ and }0\leq\phi\leq\Theta,\\
        g, &\text{otherwise},
    \end{cases}
\end{equation}
where $G_0\approx 2.2846$, and $g\approx 0$ represents the antenna gain outside the beamwidth.

In the considered scenario, the GTs are positioned in outdoor environments, and the communication channels between the UAV and each GT are dominated by the line-of-sight (LoS) path. Consequently, the channel gain between the UAV and GT $k$ can be expressed as\footnote{For simplification, we consider the GTs are located outdoors within rural areas where the pathloss exponent is $2$, and we assume that communication via sidelobes is negligible.}
\begin{equation}
    g_k=\frac{g_0}{\left(d_k^\mathrm{UG}\right)^2},
\end{equation}
where $g_0$ denotes the channel gain at the reference distance of $1$ m.

To ensure effective communication between the UAV and the GTs, all GTs must be within the coverage area of the UAV, which means
\begin{equation}\label{eq.ca}
    \left\|\mathbf{L}^\mathrm{U}-\mathbf{L}_k^\mathrm{G}\right\|\leq H_\mathrm{U}\tan\Theta,\forall k\in\mathcal{K}.
\end{equation}
We employ frequency-division multiple access (FDMA) for UAV-to-GT communication, then the achievable downlink transmission rate for GT $k$ satisfying constraint \eqref{eq.ca} can be written as
\begin{equation}
    r_k=b_k\log_2\left(1+\frac{G_0 g_k p_k}{\Theta^2 b_k N_0}\right),
\end{equation}
where $b_k$ is the allocated bandwidth for GT $k$, and $p_k$ is the allocated transmit power for GT $k$.

\subsection{Latency and Energy Model}
Latency and energy efficiency are both important metrics in communication systems, particularly in SAGINs. SAGINs are characterized by exceptionally long communication distances, which require great attention to latency. Additionally, the resource-constrained nature of SAGINs highlights the importance of energy efficiency.

\subsubsection{Satellite to UAV}
The computation latency caused by satellite computation can be modeled as
\begin{equation}
    t_\mathrm{S}=\frac{\kappa \sum_{k=1}^K a_k^\mathrm{S}O_k(\rho_k)}{F_\mathrm{S}},
\end{equation}
where $\kappa$ is a constant, and $F_\mathrm{S}$ denotes the computation capacity of the satellite.

Then, the computation energy cost by the satellite can be written as
\begin{equation}
    e_\mathrm{S}=\tau t_\mathrm{S} F_\mathrm{S}^3,
\end{equation}
where $\tau$ is a constant.

The communication latency of the satellite-to-UAV link is comprised of transmission delay and propagation delay. The propagation delay cannot be neglected since the vast distance between the satellite and the UAV, often spanning thousands of kilometers.

The transmission delay of the satellite-to-UAV communication can be expressed as
\begin{equation}
    t_\mathrm{T}=\frac{\sum_{k=1}^K \left[a_k^\mathrm{S}C_k + \left(1-a_k^\mathrm{S}\right)D_k\right]}{r_\mathrm{SU}},
\end{equation}
which is data size over transmission rate.
The propagation delay of the satellite-to-UAV communication can be expressed as
\begin{equation}
    t_\mathrm{P}=\frac{d^\mathrm{SU}}{c},
\end{equation}
where $c$ is the speed of light.
Then, the total communication latency of the satellite-to-UAV communication can be written as
\begin{equation}
    t_\mathrm{SU}=t_\mathrm{T}+t_\mathrm{P}.
\end{equation}

Afterwards, the satellite-to-UAV communication energy can be calculated as
\begin{equation}
    e_\mathrm{SU}=t_\mathrm{T}P_\mathrm{S}.
\end{equation}

\subsubsection{UAV to GT}
The computation latency caused by UAV computation for GT $k$ can be modeled as
\begin{equation}
    t_k^\mathrm{U}=\frac{\kappa a_k^\mathrm{U}O_k(\rho_k)}{f_k},
\end{equation}
where $f_k$ denotes the computation capacity of the UAV that is allocated for GT $k$.

Then, the total computation energy cost by the UAV can be written as
\begin{equation}
    e_\mathrm{U}=\tau \sum_{k=1}^K t_k^\mathrm{U} f_k^3.
\end{equation}

As the distance between the UAV and each GT is only a few hundred meters, the propagation delay in UAV-to-GT communication can be neglected.

Then, the communication latency from the UAV to GT $k$ that is in coverage can be expressed as
\begin{equation}
    t_k^\mathrm{UG}=\frac{\left(a_k^\mathrm{S}+a_k^\mathrm{U}\right)\rho_k D_k+\left(1-a_k^\mathrm{S}-a_k^\mathrm{U}\right)D_k}{r_k}.
\end{equation}

Afterwards, the total UAV-to-GT communication energy can be calculated as
\begin{equation}
    e_\mathrm{UG}=\sum_{k=1}^K t_k^\mathrm{UG}p_k.
\end{equation}

\begin{figure}[t]
\centering
\includegraphics[width=\linewidth]{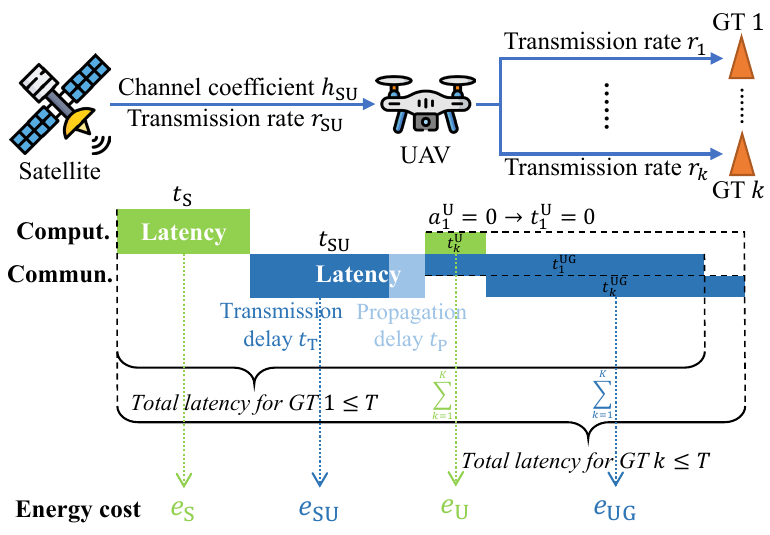}
\caption{The framework of the considered SAGIN-enabled PSC network.}
\label{fg.f}
\end{figure}

The general framework of the considered SAGIN-enabled PSC network is illustrated in Fig.~\ref{fg.f}.

\subsection{Problem Formulation}
Given the defined system model, our goal is to minimize the energy consumption of the SAGIN-enabled PSC network while considering latency requirement, power budget and suitable location of the UAV, bandwidth and computation capacity allocation of the UAV, semantic compression ratio for each GT, and computation task allocation between the satellite and the UAV. Mathematically, the energy minimization problem can be formulated as
\begin{subequations}\label{eq.pf}
    \begin{align}
        \min_{\mathbf{L}^\mathrm{U},H_\mathrm{U},\Theta,\mathbf{b},\mathbf{f},\mathbf{p},\bm{\rho},\mathbf{a}^\mathrm{S},\mathbf{a}^\mathrm{U}} \quad & e_\mathrm{S}+e_\mathrm{SU}+e_\mathrm{U}+e_\mathrm{UG}, \tag{\ref{eq.pf}}\\
        \textrm{s.t.} \hspace{0.65in} &\hspace{-0.4in} t_\mathrm{S}+t_\mathrm{SU}+t_k^\mathrm{U}+t_k^\mathrm{UG}\leq T,\forall k\in\mathcal{K},\label{eq.c1}\\
        &\hspace{-0.4in} \sum_{k=1}^K p_k\leq P_\mathrm{U},\label{eq.c2}\\
        &\hspace{-0.4in} \left\|\mathbf{L}^\mathrm{U}-\mathbf{L}_k^\mathrm{G}\right\|\leq H_\mathrm{U}\tan\Theta,\forall k\in\mathcal{K},\label{eq.c3}\\
        &\hspace{-0.4in} H_\mathrm{U}^{\min}\leq H_\mathrm{U}\leq H_\mathrm{U}^{\max},\label{eq.c4}\\
        &\hspace{-0.4in} \sum_{k=1}^K b_k\leq B_\mathrm{U},\label{eq.c5}\\
        &\hspace{-0.4in} \sum_{k=1}^K f_k\leq F_\mathrm{U},\label{eq.c6}\\
        &\hspace{-0.4in} \rho_{k}^{\min}\leq\rho_k\leq 1,\forall k\in\mathcal{K},\label{eq.c7}\\
        &\hspace{-0.4in} a_k^\mathrm{S}+a_k^\mathrm{U}\leq 1,\forall k\in\mathcal{K},\label{eq.c8}\\
        &\hspace{-0.4in} a_k^\mathrm{S},a_k^\mathrm{U}\in \left\{0,1\right\},\forall k\in\mathcal{K},\label{eq.c9}\\
        &\hspace{-0.4in} \Theta_{\min}\leq\Theta\leq\Theta_{\max},\label{eq.c10}\\
        &\hspace{-0.4in} b_k,f_k,p_k\geq 0,\forall k\in\mathcal{K},\label{eq.c11}
    \end{align}
\end{subequations}
where $\mathbf{b}=\left[b_1,\cdots,b_K\right]^\mathrm{T}$, $\mathbf{f}=\left[f_1,\cdots,f_K\right]^\mathrm{T}$, $\mathbf{p}=\left[p_1,\cdots,p_K\right]^\mathrm{T}$, $\bm{\rho}=\left[\rho_1,\cdots,\rho_K\right]^\mathrm{T}$, $\mathbf{a}^\mathrm{S}=\left[a^\mathrm{S}_1,\cdots,a^\mathrm{S}_K\right]^\mathrm{T}$, and $\mathbf{a}^\mathrm{U}=\left[a^\mathrm{U}_1,\cdots,a^\mathrm{U}_K\right]^\mathrm{T}$. $T$ is the maximum tolerable latency of each GT, $P_\mathrm{U}$ is the total power budget of the UAV, $\left[H_\mathrm{U}^{\min},H_\mathrm{U}^{\max}\right]$ is the feasible altitude range of the UAV, constrained by both obstacle heights and regulatory limitations, $B_\mathrm{U}$ is the total bandwidth of the UAV, $F_\mathrm{U}$ is the total computation capacity of the UAV, $\rho_{k}^{\min}$ is the minimum achievable semantic compression ratio of GT $k$, and $\left[\Theta_{\min},\Theta_{\max}\right]$ is the feasible range of half-beamwidth for the UAV's antenna, as determined by practical antenna beamwidth tuning techniques.

In problem \eqref{eq.pf}, constraint \eqref{eq.c1} requires that the SAGIN-enabled PSC network cannot have a latency exceeding $T$ for all GTs, which necessitates a careful trade-off between communication and computation. Constraints \eqref{eq.c2}, \eqref{eq.c5}, and \eqref{eq.c6} impose limitations on the total power, bandwidth, and computation capacity resources allocated for each GT. Constraints \eqref{eq.c3}, \eqref{eq.c4}, and \eqref{eq.c10} limit the location of the UAV and the half-beamwidth of its antenna. Constraint \eqref{eq.c7} determines the range of the semantic compression ratio of each GT. Constraints \eqref{eq.c8} and \eqref{eq.c9} govern the allocation of computation tasks between the satellite and the UAV. Finally, constraint \eqref{eq.c11} guarantees the non-negativity of bandwidth, computation capacity, and transmit power.

It is generally hard to solve problem \eqref{eq.pf} since both the objective function and constraint \eqref{eq.c1} are non-convex. Additionally, the integer constraint \eqref{eq.c9} and the presence of the piecewise function $O_k(\rho_k)$ further complicate the optimization process. To address these challenges and achieve a polynomial-time solution for problem \eqref{eq.pf}, we propose an iterative algorithm that leverages the alternating method.

\section{Algorithm Design}\label{ad}
In this section, we propose an alternating algorithm to iteratively solve problem \eqref{eq.pf} by optimizing six subproblems: satellite-UAV computation task allocation, semantic compression ratio optimization, optimal computation capacity allocation, optimal power and bandwidth allocation, optimal altitude and beamwidth, and optimal location planning.

\subsection{Satellite-UAV Computation Task Allocation}
With given semantic compression ratio, computation capacity, power, and bandwidth allocation, altitude, beamwidth, and location planning, problem \eqref{eq.pf} can be simplified as
\begin{subequations}\label{eq.sucta}
    \begin{align}
        \min_{\mathbf{a}^\mathrm{S},\mathbf{a}^\mathrm{U}} \quad & \kappa \tau \sum_{k=1}^K a_k^\mathrm{S}O_k(\rho_k) F_\mathrm{S}^2\notag\\
        &+\frac{P_\mathrm{S} \sum_{k=1}^K \left[a_k^\mathrm{S}\rho_k D_k + \left(1-a_k^\mathrm{S}\right)D_k\right]}{r_\mathrm{SU}}\notag\\
        &+\kappa\tau \sum_{k=1}^K a_k^\mathrm{U}O_k(\rho_k)f_k^2\notag\\
        &+\sum_{k=1}^K \frac{p_k D_k\left[\left(a_k^\mathrm{S}+a_k^\mathrm{U}\right)\rho_k+\left(1-a_k^\mathrm{S}-a_k^\mathrm{U}\right)\right]}{r_k},\tag{\ref{eq.sucta}}\\
        \textrm{s.t.} \hspace{0.2in} & \frac{\kappa \sum_{k=1}^K a_k^\mathrm{S}O_k(\rho_k)}{F_\mathrm{S}}\notag\\
        &+\frac{\sum_{k=1}^K \left[a_k^\mathrm{S}\rho_k D_k + \left(1-a_k^\mathrm{S}\right)D_k\right]}{r_\mathrm{SU}}+\frac{d^\mathrm{SU}}{c}\notag\\
        &+\frac{\kappa a_k^\mathrm{U}O_k(\rho_k)}{f_k}\notag\\
        &+D_k\frac{\left(a_k^\mathrm{S}+a_k^\mathrm{U}\right)\rho_k+\left(1-a_k^\mathrm{S}-a_k^\mathrm{U}\right)}{r_k}\leq T,\forall k\in\mathcal{K},\label{eq.su.c1}\\
        & a_k^\mathrm{S}+a_k^\mathrm{U}\leq 1,\forall k\in\mathcal{K},\label{eq.su.c2}\\
        & a_k^\mathrm{S},a_k^\mathrm{U}\in \left\{0,1\right\},\forall k\in\mathcal{K}.\label{eq.su.c3}
    \end{align}
\end{subequations}
The difficulty in solving problem \eqref{eq.sucta} arises from the discrete nature of the value space for $\mathbf{a}^\mathrm{S}$ and $\mathbf{a}^\mathrm{U}$. This characteristic transforms problem \eqref{eq.sucta} into a discrete optimization problem, whose complexity of finding the optimal solution is often significantly high.


To deal with the discrete difficulty of problem \eqref{eq.sucta}, we first relax the integer constraint \eqref{eq.su.c2} with
\begin{equation}\label{eq.ir}
    a_k^\mathrm{S},a_k^\mathrm{U}\in \left[0,1\right],\forall k\in\mathcal{K}.
\end{equation}
Then, problem \eqref{eq.sucta} becomes a convex optimization problem which can be addressed by the dual method.

The dual problem of problem \eqref{eq.sucta} after integer relaxation \eqref{eq.ir} can be written as
\begin{equation}
    \max_{\bm{\lambda}} \quad D\left(\bm{\lambda}\right),
\end{equation}
where
\begin{equation}\label{eq.dl}
    D\left(\bm{\lambda}\right)=\left\{
        \begin{aligned}
        \min_{\mathbf{a}^\mathrm{S},\mathbf{a}^\mathrm{U}} \quad & L\left(\mathbf{a}^\mathrm{S},\mathbf{a}^\mathrm{U},\bm{\lambda}\right),\\
        \textrm{s.t.} \hspace{0.2in} & a_k^\mathrm{S}+a_k^\mathrm{U}\leq 1,\forall k\in\mathcal{K},\\
        & a_k^\mathrm{S},a_k^\mathrm{U}\in \left[0,1\right],\forall k\in\mathcal{K},
    \end{aligned}
    \right.
\end{equation}
with
\begin{align}
    L\left(\mathbf{a}^\mathrm{S},\mathbf{a}^\mathrm{U},\bm{\lambda}\right)=&\kappa \tau \sum_{k=1}^K a_k^\mathrm{S}O_k(\rho_k) F_\mathrm{S}^2\notag\\
        &+\frac{P_\mathrm{S} \sum_{k=1}^K \left[a_k^\mathrm{S}\rho_k D_k + \left(1-a_k^\mathrm{S}\right)D_k\right]}{r_\mathrm{SU}}\notag\\
        &+\kappa\tau \sum_{k=1}^K a_k^\mathrm{U}O_k(\rho_k)f_k^2\notag\\
        &+\sum_{k=1}^K p_k D_k\frac{\left(a_k^\mathrm{S}+a_k^\mathrm{U}\right)\rho_k+\left(1-a_k^\mathrm{S}-a_k^\mathrm{U}\right)}{r_k}\notag\\
        &+\sum_{k=1}^K \lambda_k \Bigg( \frac{\kappa \sum_{k=1}^K a_k^\mathrm{S}O_k(\rho_k)}{F_\mathrm{S}}\notag\\
        &+\frac{\sum_{k=1}^K \left[a_k^\mathrm{S}\rho_k D_k + \left(1-a_k^\mathrm{S}\right)D_k\right]}{r_\mathrm{SU}}+\frac{d^\mathrm{SU}}{c}\notag\\
        &+\frac{\kappa a_k^\mathrm{U}O_k(\rho_k)}{f_k}\notag\\
        &+D_k\frac{\left(a_k^\mathrm{S}+a_k^\mathrm{U}\right)\rho_k+\left(1-a_k^\mathrm{S}-a_k^\mathrm{U}\right)}{r_k}-T\Bigg),
\end{align}
and $\bm{\lambda}=\left[\lambda_1,\cdots,\lambda_K\right]$ is non-negative Lagrange multiplier vector with respect to the corresponding constraint \eqref{eq.su.c1}.

The objective function in \eqref{eq.dl} is linear, we can write the coefficient corresponding to $a_k^\mathrm{S}$ as
\begin{align}
    A_k^\mathrm{S}=&\kappa\tau O_k(\rho_k)F_\mathrm{S}^2- D_k(1-\rho_k)\left(\frac{P_\mathrm{S}}{r_\mathrm{SU}}+\frac{p_k}{r_k}\right)\notag\\
    &+\lambda_k\left[\frac{\kappa O_k(\rho_k)}{F_\mathrm{S}}-D_k(1-\rho_k)\left(\frac{1}{r_\mathrm{SU}}+\frac{1}{r_k}\right)\right],
\end{align}
and write the coefficient corresponding to $a_k^\mathrm{U}$ as
\begin{align}
    A_k^\mathrm{U}=&\kappa\tau O_k(\rho_k)f_k^2- D_k (1-\rho_k)\frac{p_k}{r_k}\notag\\
    &+\lambda_k\left[\frac{\kappa O_k(\rho_k)}{f_k}-D_k(1-\rho_k)\frac{1}{r_k}\right].
\end{align}

Considering constraint $a_k^\mathrm{S}+a_k^\mathrm{U}\leq 1$, we can obtain the optimal solution as
\begin{equation}\label{eq.aa}
    \left(a_k^\mathrm{S},a_k^\mathrm{U}\right)^*=
    \begin{cases}
        (0,0),&\text{if }\left(A_k^\mathrm{S}\geq 0\text{ and }A_k^\mathrm{U}\geq 0\right),\\
        (0,1),&\text{if }\left(A_k^\mathrm{S}\geq 0\text{ and }A_k^\mathrm{U}<0\right)\\
        &\hspace{0.5in}\text{ or }\left(A_k^\mathrm{U}<A_k^\mathrm{S}<0\right),\\
        (1,0),&\text{if }\left(A_k^\mathrm{S}<0\text{ and }A_k^\mathrm{U}\geq 0\right)\\
        &\hspace{0.5in}\text{ or }\left(A_k^\mathrm{S}<A_k^\mathrm{U}<0\right),
    \end{cases}
\end{equation}

The value of $\bm{\lambda}$ can be determined by the sub-gradient method, and the updating process can be given by
\begin{align}\label{eq.ul}
    \lambda_k=\Bigg[\lambda_k+&\xi \Bigg(\frac{\kappa \sum_{k=1}^K a_k^\mathrm{S}O_k(\rho_k)}{F_\mathrm{S}}\notag\\
        &+\frac{\sum_{k=1}^K \left[a_k^\mathrm{S}\rho_k D_k + \left(1-a_k^\mathrm{S}\right)D_k\right]}{r_\mathrm{SU}}+\frac{d^\mathrm{SU}}{c}\notag\\
        &+\frac{\kappa a_k^\mathrm{U}O_k(\rho_k)}{f_k}\notag\\
        &+D_k\frac{\left(a_k^\mathrm{S}+a_k^\mathrm{U}\right)\rho_k+\left(1-a_k^\mathrm{S}-a_k^\mathrm{U}\right)}{r_k}-T\Bigg)\Bigg]^+,
\end{align}
where $[x]^+=\max\{x,0\}$, and $\xi>0$ is the dynamic step size \cite{bertsekas2009convex}. By iteratively updating $\left(a_k^\mathrm{S},a_k^\mathrm{U}\right)$ according to \eqref{eq.aa} and $\lambda_k$ according to \eqref{eq.ul}, we can obtain the optimal solution of problem \eqref{eq.sucta} with zero duality gap. Note that although we relaxed the integer constraint \eqref{eq.su.c3} to be continuous, the optimal solution we derived always satisfies the discrete constraint $a_k^\mathrm{S},a_k^\mathrm{U}\in \left\{0,1\right\}$ in accordance with \eqref{eq.aa}. Therefore, the integer relaxation does not affect the optimality of problem \eqref{eq.sucta}.

\subsection{Semantic Compression Ratio Optimization}
With given satellite-UAV computation task allocation, computation capacity, power, and bandwidth allocation, altitude, beamwidth, and location planning, problem \eqref{eq.pf} can be simplified as
\begin{subequations}\label{eq.scro}
    \begin{align}
        \min_{\bm{\rho}} \quad &  \kappa \tau \sum_{k=1}^K a_k^\mathrm{S}O_k(\rho_k) F_\mathrm{S}^2\notag\\
        &+\frac{P_\mathrm{S} \sum_{k=1}^K \left[a_k^\mathrm{S}\rho_k D_k + \left(1-a_k^\mathrm{S}\right)D_k\right]}{r_\mathrm{SU}}\notag\\
        &+\kappa\tau \sum_{k=1}^K a_k^\mathrm{U}O_k(\rho_k)f_k^2\notag\\
        &+\sum_{k=1}^K \frac{p_k D_k\left[\left(a_k^\mathrm{S}+a_k^\mathrm{U}\right)\rho_k+\left(1-a_k^\mathrm{S}-a_k^\mathrm{U}\right)\right]}{r_k},\tag{\ref{eq.scro}}\\
        \textrm{s.t.} \quad & \frac{\kappa \sum_{k=1}^K a_k^\mathrm{S}O_k(\rho_k)}{F_\mathrm{S}}\notag\\
        &+\frac{\sum_{k=1}^K \left[a_k^\mathrm{S}\rho_k D_k + \left(1-a_k^\mathrm{S}\right)D_k\right]}{r_\mathrm{SU}}+\frac{d^\mathrm{SU}}{c}\notag\\
        &+\frac{\kappa a_k^\mathrm{U}O_k(\rho_k)}{f_k}\notag\\
        &+D_k\frac{\left(a_k^\mathrm{S}+a_k^\mathrm{U}\right)\rho_k+\left(1-a_k^\mathrm{S}-a_k^\mathrm{U}\right)}{r_k}\leq T,\forall k\in\mathcal{K},\label{eq.scro.c1}\\
        & \rho_k^{\min}\leq\rho_k\leq 1,\forall k\in\mathcal{K}\label{eq.scro.c2}.
    \end{align}
\end{subequations}
The difficulty in solving problem \eqref{eq.scro} lies in the piecewise function $O_k(\rho_k)$, which leads to non-smooth optimization.

To address this difficulty, we suggest using the binary variable $\alpha_{kd}\in \{0,1\}$ to signify the linear segment level of $O_k(\rho_k)$. When $\alpha_{kd} =1$, the computation overhead function $O_k(\rho_k)$ is associated with the $d$-th segment. Thus, it can be represented as $O_k(\rho_k)=A_{kd}\rho_k+B_{kd}$. Conversely, if $\alpha_{kd}=0$, the computation overhead function does not pertain to the $d$-th segment. By introducing the binary variable $\alpha_{kd}$, we can rewrite the computation overhead function as
\begin{equation}
    O_k(\rho_k)=\sum_{d=1}^{D} \alpha_{kd}\left(A_{kd}\rho_k+B_{kd}\right),
\end{equation}
where $D$ represents the total number of segments of the piecewise function. Furthermore, we have
\begin{equation}
    \sum_{d=1}^{D}\alpha_{kd}=1,\alpha_{kd}\in \{0,1\},
\end{equation}
since each GT corresponds to only one linear segment level.

After the above reformulation, we first aim to roughly determine the segment of the piecewise function. To achieve this, we use the midpoint of each segment $d$ to approximate the value of this segment, which can be given by
\begin{equation}
    \rho_{kd}=\frac{C_{kd}+C_{k(d-1)}}{2},\forall d\in\mathcal{D}^\mathrm{s},\forall k\in\mathcal{K},
\end{equation}
with $C_{k0}=1,\forall k\in\mathcal{K}$.

Then, the segment selection problem can be expressed as
\begin{subequations}\label{eq.ss}
    \begin{align}
        \min_{\bm{\alpha}} \quad &  \kappa \tau F_\mathrm{S}^2 \sum_{k=1}^K \sum_{d=1}^{D} a_k^\mathrm{S} \alpha_{kd}\left(A_{kd}\rho_{kd}+B_{kd}\right)\notag\\
        &+\frac{P_\mathrm{S} \sum_{k=1}^K D_k \left[a_k^\mathrm{S} \left(\sum_{d=1}^{D}\alpha_{kd}\rho_{kd}\right) + \left(1-a_k^\mathrm{S}\right)\right]}{r_\mathrm{SU}}\notag\\
        &+\kappa\tau \sum_{k=1}^K \sum_{d=1}^{D} a_k^\mathrm{U} f_k^2 \alpha_{kd}\left(A_{kd}\rho_{kd}+B_{kd}\right)\notag\\
        &+\sum_{k=1}^K p_k D_k \frac{1-a_k^\mathrm{S}-a_k^\mathrm{U}+\left(a_k^\mathrm{S}+a_k^\mathrm{U}\right)\sum_{d=1}^{D}\alpha_{kd}\rho_{kd}}{r_k},\tag{\ref{eq.ss}}\\
        \textrm{s.t.} \quad & \frac{\kappa \sum_{k=1}^K \sum_{d=1}^{D} a_k^\mathrm{S} \alpha_{kd}\left(A_{kd}\rho_{kd}+B_{kd}\right)}{F_\mathrm{S}}\notag\\
        &+\frac{\sum_{k=1}^K D_k \left[a_k^\mathrm{S} \left(\sum_{d=1}^{D}\alpha_{kd}\rho_{kd}\right) + 1-a_k^\mathrm{S}\right]}{r_\mathrm{SU}}+\frac{d^\mathrm{SU}}{c}\notag\\
        &+\frac{\kappa a_k^\mathrm{U} \sum_{d=1}^{D} \alpha_{kd}\left(A_{kd}\rho_{kd}+B_{kd}\right)}{f_k}\notag\\
        &+D_k \frac{1-a_k^\mathrm{S}-a_k^\mathrm{U}+\left(a_k^\mathrm{S}+a_k^\mathrm{U}\right)\sum_{d=1}^{D}\alpha_{kd}\rho_{kd}}{r_k}\leq T,\notag\\
        &\hspace{2.2in}\forall k\in\mathcal{K},\label{eq.ss.c1}\\
        & \sum_{d=1}^{D}\alpha_{kd}=1,\forall k\in\mathcal{K},\label{eq.ss.c2}\\
        & \alpha_{kd}\in \{0,1\},\forall d\in\mathcal{D}^\mathrm{s},\forall k\in\mathcal{K}.\label{eq.ss.c3}
    \end{align}
\end{subequations}
where $\bm{\alpha}=\left[\bm{\alpha}_{1},\cdots,\bm{\alpha}_{k},\cdots,\bm{\alpha}_{K}\right]$ with $\bm{\alpha}_{k}=[\alpha_{k1};\cdots;$ $\alpha_{kD}]$.

Similarly, we relax the integer constraint \eqref{eq.ss.c3} with
\begin{equation}\label{eq.ssir}
    \alpha_{kd}\in [0,1],\forall d\in\mathcal{D}^\mathrm{s},\forall k\in\mathcal{K}.
\end{equation}
Then, problem \eqref{eq.ss} becomes a convex optimization problem which can be tackled by the dual method.

The dual problem of problem \eqref{eq.ss} after integer relaxation \eqref{eq.ssir} can be written as
\begin{equation}
    \max_{\bm{\gamma}} \quad \hat{D}\left(\bm{\gamma}\right),
\end{equation}
where
\begin{equation}\label{eq.ssdl}
    \hat{D}\left(\bm{\gamma}\right)=\left\{
        \begin{aligned}
        \min_{\bm{\alpha}} \quad & \hat{L}\left(\bm{\alpha},\bm{\gamma}\right),\\
        \textrm{s.t.} \quad & \sum_{d=1}^{D}\alpha_{kd}=1,\forall k\in\mathcal{K},\\
        & \alpha_{kd}\in [0,1],\forall d\in\mathcal{D}^\mathrm{s},\forall k\in\mathcal{K},
    \end{aligned}
    \right.
\end{equation}
with
\begin{align}
    \hat{L}(\bm{\alpha},&\bm{\gamma})=\kappa \tau F_\mathrm{S}^2 \sum_{k=1}^K \sum_{d=1}^{D} a_k^\mathrm{S} \alpha_{kd}\left(A_{kd}\rho_{kd}+B_{kd}\right)\notag\\
        &+\frac{P_\mathrm{S} \sum_{k=1}^K D_k \left[a_k^\mathrm{S} \left(\sum_{d=1}^{D}\alpha_{kd}\rho_{kd}\right) + \left(1-a_k^\mathrm{S}\right)\right]}{r_\mathrm{SU}}\notag\\
        &+\kappa\tau \sum_{k=1}^K \sum_{d=1}^{D} a_k^\mathrm{U} f_k^2 \alpha_{kd}\left(A_{kd}\rho_{kd}+B_{kd}\right)\notag\\
        &+\sum_{k=1}^K p_k D_k \frac{1-a_k^\mathrm{S}-a_k^\mathrm{U}+\left(a_k^\mathrm{S}+a_k^\mathrm{U}\right)\sum_{d=1}^{D}\alpha_{kd}\rho_{kd}}{r_k}\notag\\
        &+\sum_{k=1}^K \gamma_k \Bigg( \frac{\kappa \sum_{k=1}^K \sum_{d=1}^{D} a_k^\mathrm{S} \alpha_{kd}\left(A_{kd}\rho_{kd}+B_{kd}\right)}{F_\mathrm{S}}\notag\\
        &+\frac{\sum_{k=1}^K D_k \left[a_k^\mathrm{S} \left(\sum_{d=1}^{D}\alpha_{kd}\rho_{kd}\right) + 1-a_k^\mathrm{S}\right]}{r_\mathrm{SU}}+\frac{d^\mathrm{SU}}{c}\notag\\
        &+\frac{\kappa a_k^\mathrm{U} \sum_{d=1}^{D} \alpha_{kd}\left(A_{kd}\rho_{kd}+B_{kd}\right)}{f_k}\notag\\
        &+D_k \frac{1-a_k^\mathrm{S}-a_k^\mathrm{U}+\left(a_k^\mathrm{S}+a_k^\mathrm{U}\right)\sum_{d=1}^{D}\alpha_{kd}\rho_{kd}}{r_k}-T\Bigg),
\end{align}
and $\bm{\gamma}=\left[\gamma_1,\cdots,\gamma_K\right]$ is non-negative Lagrange multiplier vector with respect to the corresponding constraint \eqref{eq.ss.c1}.

The objective function in \eqref{eq.ssdl} is linear, we can write the coefficient corresponding to $\alpha_{kd}$ as
\begin{align}
    S_{kd}=&\kappa \tau F_\mathrm{S}^2 a_k^\mathrm{S} \left(A_{kd}\rho_{kd}+B_{kd}\right) + \frac{P_\mathrm{S} D_k a_k^\mathrm{S} \rho_{kd}}{r_\mathrm{SU}}\notag\\
    &+ \kappa \tau f_k^2 a_k^\mathrm{U} \left(A_{kd}\rho_{kd}+B_{kd}\right) + \frac{p_k D_k \rho_{kd} \left(a_k^\mathrm{S}+a_k^\mathrm{U}\right)}{r_k}\notag\\
    &+\gamma_k\Bigg[\frac{\kappa a_k^\mathrm{S} \left(A_{kd}\rho_{kd}+B_{kd}\right)}{F_\mathrm{S}} + \frac{D_k a_k^\mathrm{S} \rho_{kd}}{r_\mathrm{SU}}\notag\\
    &+ \frac{\kappa a_k^\mathrm{U} \left(A_{kd}\rho_{kd}+B_{kd}\right)}{f_k} + \frac{D_k \rho_{kd} \left(a_k^\mathrm{S}+a_k^\mathrm{U}\right)}{r_k}\Bigg].
\end{align}

Due to constraint $\sum_{d=1}^{D}\alpha_{kd}=1,\forall k\in\mathcal{K}$ and the linear objective function, we let $\alpha_{kd}$ corresponding to the smallest coefficient be $1$ for any $k$. Hence, the optimal segment selection can be given by
\begin{equation}\label{eq.oa}
    \alpha_{kd}^*=
    \begin{cases}
        1,&\text{if }d=\arg\min_{d\in\mathcal{D}^\mathrm{s}}S_{kd},\\
        0,&\text{otherwise,}
    \end{cases}
\end{equation}

The value of $\bm{\gamma}$ can be determined by the sub-gradient method, and the updating process can be given by
\begin{align}
    \gamma_k&=\Bigg[\gamma_k+\xi \Bigg( \frac{\kappa \sum_{k=1}^K \sum_{d=1}^{D} a_k^\mathrm{S} \alpha_{kd}\left(A_{kd}\rho_{kd}+B_{kd}\right)}{F_\mathrm{S}}\notag\\
        &+\frac{\sum_{k=1}^K D_k \left[a_k^\mathrm{S} \left(\sum_{d=1}^{D}\alpha_{kd}\rho_{kd}\right) + 1-a_k^\mathrm{S}\right]}{r_\mathrm{SU}}+\frac{d^\mathrm{SU}}{c}\notag\\
        &+\frac{\kappa a_k^\mathrm{U} \sum_{d=1}^{D} \alpha_{kd}\left(A_{kd}\rho_{kd}+B_{kd}\right)}{f_k}\notag\\
        &+D_k \frac{1-a_k^\mathrm{S}-a_k^\mathrm{U}+\left(a_k^\mathrm{S}+a_k^\mathrm{U}\right)\sum_{d=1}^{D}\alpha_{kd}\rho_{kd}}{r_k}-T\Bigg)\Bigg]^+,
\end{align}
By iteratively updating $\alpha_{kd}$ and $\gamma_k$, we can obtain the optimal solution of problem \eqref{eq.ss} with zero duality gap.

After solving the segment selection problem, we can determine which segment does $O_k(\rho_k)$ belong to. Denote the optimal segment of $O_k(\rho_k)$ by $d_k^*$, we can rewrite problem \eqref{eq.scro} as
\begin{subequations}\label{eq.scroass}
    \begin{align}
        \min_{\bm{\rho}} \quad &  \kappa \tau \sum_{k=1}^K a_k^\mathrm{S}\left(A_{kd_k^*}\rho_k+B_{kd_k^*}\right) F_\mathrm{S}^2\notag\\
        &+\frac{P_\mathrm{S} \sum_{k=1}^K \left[a_k^\mathrm{S}\rho_k D_k + \left(1-a_k^\mathrm{S}\right)D_k\right]}{r_\mathrm{SU}}\notag\\
        &+\kappa\tau \sum_{k=1}^K a_k^\mathrm{U}\left(A_{kd_k^*}\rho_k+B_{kd_k^*}\right)f_k^2\notag\\
        &+\sum_{k=1}^K \frac{p_k D_k\left[\left(a_k^\mathrm{S}+a_k^\mathrm{U}\right)\rho_k+\left(1-a_k^\mathrm{S}-a_k^\mathrm{U}\right)\right]}{r_k},\tag{\ref{eq.scroass}}\\
        \textrm{s.t.} \quad & \frac{\kappa \sum_{k=1}^K a_k^\mathrm{S}\left(A_{kd_k^*}\rho_k+B_{kd_k^*}\right)}{F_\mathrm{S}}\notag\\
        &+\frac{\sum_{k=1}^K \left[a_k^\mathrm{S}\rho_k D_k + \left(1-a_k^\mathrm{S}\right)D_k\right]}{r_\mathrm{SU}}+\frac{d^\mathrm{SU}}{c}\notag\\
        &+\frac{\kappa a_k^\mathrm{U}\left(A_{kd_k^*}\rho_k+B_{kd_k^*}\right)}{f_k}\notag\\
        &+D_k\frac{\left(a_k^\mathrm{S}+a_k^\mathrm{U}\right)\rho_k+\left(1-a_k^\mathrm{S}-a_k^\mathrm{U}\right)}{r_k}\leq T,\forall k\in\mathcal{K},\\
        & C_{k d_k^*}\leq\rho_k\leq C_{k (d_k^*-1)},\forall k\in\mathcal{K},
    \end{align}
\end{subequations}
which is a linear optimization problem and can be addressed using existing toolbox.

\subsection{Optimal Computation Capacity Allocation}
With given satellite-UAV computation task allocation, semantic compression ratio, power, and bandwidth allocation, altitude, beamwidth, and location planning, problem \eqref{eq.pf} can be simplified as
\begin{subequations}\label{eq.cca}
    \begin{align}
        \min_{\mathbf{f}} \quad & \sum_{k=1}^K a_k^\mathrm{U}O_k(\rho_k)f_k^2,\tag{\ref{eq.cca}}\\
        \textrm{s.t.} \quad &t_\mathrm{S}+t_\mathrm{SU}+\frac{\kappa a_k^\mathrm{U}O_k(\rho_k)}{f_k}+t_k^\mathrm{UG}\leq T,\forall k\in\mathcal{K},\label{eq.cca.c1}\\
        & \sum_{k=1}^K f_k\leq F_\mathrm{U},\label{eq.cca.c2}\\
        &\ f_k\geq 0,\forall k\in\mathcal{K}.\label{eq.cca.c3}
    \end{align}
\end{subequations}

To solve problem \eqref{eq.cca}, we obtain the following theorem.
\begin{theorem}
The optimal solution of problem \eqref{eq.cca} is
\begin{equation}
    f_k=\frac{\kappa a_k^\mathrm{U}O_k(\rho_k)}{T-t_\mathrm{S}-t_\mathrm{SU}-t_k^\mathrm{UG}},\forall k\in\mathcal{K}.
\end{equation}
\end{theorem}

\begin{IEEEproof}
For those GTs with $a_k^\mathrm{U}=0$, we can simply set $f_k=0$ because the UAV does not need to compute for these GTs. For other GTs with $a_k^\mathrm{U}=1$, we can combine constraints \eqref{eq.cca.c1} and \eqref{eq.cca.c3} as
\begin{equation}\label{eq.cca.2c1}
    f_k\geq \frac{\kappa a_k^\mathrm{U}O_k(\rho_k)}{T-t_\mathrm{S}-t_\mathrm{SU}-t_k^\mathrm{UG}}>0,\forall k\in\mathcal{K},
\end{equation}
where $\frac{\kappa a_k^\mathrm{U}O_k(\rho_k)}{T-t_\mathrm{S}-t_\mathrm{SU}-t_k^\mathrm{UG}}$ is a constant in problem \eqref{eq.cca}.

Then, the Lagrange function of problem \eqref{eq.cca} can be given by
\begin{align}\label{eq.cca.l}
    L\left(\mathbf{f},\bm{\mu}_1,\mu_2\right)=&\sum_{k=1}^K a_k^\mathrm{U}O_k(\rho_k)f_k^2\notag\\
    &-\sum_{k=1}^K \mu_{1k} \left(f_k- \frac{\kappa a_k^\mathrm{U}O_k(\rho_k)}{T-t_\mathrm{S}-t_\mathrm{SU}-t_k^\mathrm{UG}}\right)\notag\\
    &+\mu_2\left(\sum_{k=1}^K f_k- F_\mathrm{U}\right),
\end{align}
where $\bm{\mu}_1=\left[\mu_{11},\cdots,\mu_{1K}\right]$ is the non-negative Lagrange multiplier vector associated with constraint \eqref{eq.cca.2c1}, and $\mu_2$ is the non-negative Lagrange multiplier associated with constraint \eqref{eq.cca.c2}. The first derivative of \eqref{eq.cca.l} is
\begin{equation}
    \frac{\partial L\left(\mathbf{f},\bm{\mu}_1,\mu_2\right)}{\partial f_k}=2 a_k^\mathrm{U}O_k(\rho_k)f_k- \mu_{1k} +\mu_2.
\end{equation}
Setting $\frac{\partial L\left(\mathbf{f},\bm{\mu}_1,\mu_2\right)}{\partial f_k}=0$ yields
\begin{equation}\label{eq.kkt1}
    f_k=\frac{\mu_{1k}-\mu_2}{2 a_k^\mathrm{U}O_k(\rho_k)}.
\end{equation}
According to complementary slackness, we have
\begin{equation}\label{eq.kkt2}
    \mu_{1k} \left(f_k- \frac{\kappa a_k^\mathrm{U}O_k(\rho_k)}{T-t_\mathrm{S}-t_\mathrm{SU}-t_k^\mathrm{UG}}\right)=0.
\end{equation}
To obtain the Karush-Kuhn-Tucker (KKT) point, conditions \eqref{eq.kkt1} and \eqref{eq.kkt2} must be satisfied at the same time. Moreover, due to the fact that $f_k>0$ and $\mu_{1k},\mu_2$ are non-negative, $\mu_{1k}$ must be greater than zero, which means
\begin{equation}
    f_k- \frac{\kappa a_k^\mathrm{U}O_k(\rho_k)}{T-t_\mathrm{S}-t_\mathrm{SU}-t_k^\mathrm{UG}}=0,
\end{equation}
for those GTs with $a_k^\mathrm{U}=1$. As mentioned above, we set $f_k=0$ for those GTs with $a_k^\mathrm{U}=0$. Hence, we can obtain the closed-form solution of problem \eqref{eq.cca} as
\begin{equation}
    f_k=\frac{\kappa a_k^\mathrm{U}O_k(\rho_k)}{T-t_\mathrm{S}-t_\mathrm{SU}-t_k^\mathrm{UG}},\forall k\in\mathcal{K}.
\end{equation}
This ends the proof.
\end{IEEEproof}

\subsection{Optimal Power and Bandwidth Allocation}
With given satellite-UAV computation task allocation, semantic compression ratio, computation capacity allocation, altitude, beamwidth, and location planning, problem \eqref{eq.pf} can be simplified as
\begin{subequations}\label{eq.pb}
    \begin{align}
        \min_{\mathbf{b},\mathbf{p}} \quad & \sum_{k=1}^K \frac{p_k D_k\left[\left(a_k^\mathrm{S}+a_k^\mathrm{U}\right)\rho_k+\left(1-a_k^\mathrm{S}-a_k^\mathrm{U}\right)\right]}{b_k\log_2\left(1+\frac{G_0 g_k p_k}{\Theta^2 b_k N_0}\right)}, \tag{\ref{eq.pb}}\\
        \textrm{s.t.} \quad & \frac{D_k\left[\left(a_k^\mathrm{S}+a_k^\mathrm{U}\right)\rho_k+\left(1-a_k^\mathrm{S}-a_k^\mathrm{U}\right)\right]}{b_k\log_2\left(1+\frac{G_0 g_k p_k}{\Theta^2 b_k N_0}\right)}\notag\\
        &\hspace{0.7in}\leq T-t_\mathrm{S}-t_\mathrm{SU}-t_k^\mathrm{U},\forall k\in\mathcal{K},\label{eq.pb.c1}\\
        & \sum_{k=1}^K p_k\leq P_\mathrm{U},\label{eq.pb.c2}\\
        & \sum_{k=1}^K b_k\leq B_\mathrm{U},\label{eq.pb.c3}\\
        & b_k,p_k\geq 0,\forall k\in\mathcal{K}.\label{eq.pb.c4}
    \end{align}
\end{subequations}

It is hard to solve problem \eqref{eq.pb} due to the non-convexity of the objective function. Hence, we first try to obtain the optimal condition of problem \eqref{eq.pb} and we have the following lemma.
\begin{lemma}\label{lemma1}
The optimal $\left(\mathbf{b}^*,\mathbf{p}^*\right)$ of problem \eqref{eq.pb} satisfies
\begin{equation}\label{eq.l1.c1}
    D_k\frac{\left(a_k^\mathrm{S}+a_k^\mathrm{U}\right)\rho_k+1-a_k^\mathrm{S}-a_k^\mathrm{U}}{b_k^*\log_2\left(1+\frac{G_0 g_k p_k^*}{\Theta^2 b_k^* N_0}\right)}= T-t_\mathrm{S}-t_\mathrm{SU}-t_k^\mathrm{U},\forall k\in\mathcal{K}.
\end{equation}
\end{lemma}

\begin{IEEEproof}
Lemma \ref{lemma1} can be proved by the contradiction method. Define function
\begin{equation}
    f(x)=\frac{x}{\log_2(1+x)},x>0,
\end{equation}
whose derivative is
\begin{equation}
    f'(x)=\frac{\left(\ln 2\right) (1+x)\log_2(1+x)-x}{\left(\ln 2\right)\left[\log_2(1+x)\right]^2(1+x)},x>0.
\end{equation}
Then, define function
\begin{equation}
    g(x)=\left(\ln 2\right) (1+x)\log_2(1+x)-x,x>0,
\end{equation}
whose derivative is
\begin{equation}
    g'(x)=\left(\ln 2\right)\log_2(1+x),x>0.
\end{equation}
Obviously, $g'(x)>0$ for $x>0$. Since $g(0)=0$, we have $g(x)>0$ for $x>0$. Furthermore, since the denominator of $f'(x)$ is greater than zero for $x>0$, we have $f'(x)>0$ for $x>0$. Thus, $f(x)$ is monotonically increasing on $x>0$. Hence, with given $b_k$, the objective function of problem \eqref{eq.pb} increases with growing $p_k$.

Assume $\left(\mathbf{b},\mathbf{p}\right)$ is a feasible solution of problem \eqref{eq.pb}, if there exists one $k$ whose corresponding constraint \eqref{eq.pb.c1} holds with inequality, we can always decrease $p_k$ to obtain a smaller objective value. Therefore, for optimal $\left(\mathbf{b}^*,\mathbf{p}^*\right)$, constraint \eqref{eq.pb.c1} must hold with equality.
\end{IEEEproof}


According to lemma \ref{lemma1}, we can separate $p_k^*$ with
\begin{equation}\label{eq.p2b}
    p_k^*=\frac{b_k^* \left(2^{\frac{U_k}{b_k^*}}-1\right)}{V_k},\forall k\in\mathcal{K},
\end{equation}
where
\begin{equation}
    U_k=\frac{D_k\left[\left(a_k^\mathrm{S}+a_k^\mathrm{U}\right)\rho_k+\left(1-a_k^\mathrm{S}-a_k^\mathrm{U}\right)\right]}{T-t_\mathrm{S}-t_\mathrm{SU}-t_k^\mathrm{U}},
\end{equation}
and
\begin{equation}
    V_k=\frac{G_0 g_k}{\Theta^2 N_0},
\end{equation}
are both constants in problem \eqref{eq.pb}.

Then, problem \eqref{eq.pb} can be reformulated as
\begin{subequations}\label{eq.b}
    \begin{align}
        \min_{\mathbf{b}} \quad & \sum_{k=1}^K \frac{T-t_\mathrm{S}-t_\mathrm{SU}-t_k^\mathrm{U}}{V_k} b_k \left(2^{\frac{U_k}{b_k}}-1\right), \tag{\ref{eq.b}}\\
        \textrm{s.t.} \quad & \sum_{k=1}^K \frac{b_k \left(2^{\frac{U_k}{b_k}}-1\right)}{V_k}\leq P_\mathrm{U},\label{eq.b.c1}\\
        & \sum_{k=1}^K b_k\leq B_\mathrm{U},\label{eq.b.c2}\\
        & b_k\geq 0,\forall k\in\mathcal{K}.\label{eq.b.c3}
    \end{align}
\end{subequations}

To solve problem \eqref{eq.b}, we have the following theorem.
\begin{theorem}\label{theorem2}
Problem \eqref{eq.b} is a convex optimization problem.
\end{theorem}

\begin{IEEEproof}
Define function
\begin{equation}
    q(x)=x\left(2^\frac{a}{x}-1\right), x>0,
\end{equation}
where $a$ is a positive constant. Then, we can write its derivative as
\begin{equation}
    q'(x)=2^\frac{a}{x}-1-\frac{a(\ln 2)2^\frac{a}{x}}{x}, x>0.
\end{equation}
Furthermore,
\begin{equation}
    q''(x)=\frac{a^2(\ln 2)^2 2^\frac{a}{x}}{x^3}, x>0.
\end{equation}
Obviously, $q''(x)>0$ on $x>0$. Thus, $q(x)$ is a convex function. Therefore, the objective function of problem \eqref{eq.b} and constraint \eqref{eq.b.c1} are both convex. Since constraints \eqref{eq.b.c2} and \eqref{eq.b.c3} are also convex, problem \eqref{eq.b} is a convex optimization problem.
\end{IEEEproof}

Following theorem \ref{theorem2}, problem \eqref{eq.b} can be efficiently solved using existing convex optimization toolbox.

\subsection{Optimal Altitude and Beamwidth}
With given satellite-UAV computation task allocation, semantic compression ratio, computation capacity, power, bandwidth allocation, and location planning, problem \eqref{eq.pf} can be simplified as
\begin{subequations}\label{eq.ab}
    \begin{align}
        \min_{H_\mathrm{U},\Theta} \quad & \sum_{k=1}^K \frac{p_k D_k\left[\left(a_k^\mathrm{S}+a_k^\mathrm{U}\right)\rho_k+\left(1-a_k^\mathrm{S}-a_k^\mathrm{U}\right)\right]}{b_k\log_2\left(1+\frac{G_0 g_0 p_k}{\Theta^2 \left(\left\|\mathbf{L}^\mathrm{U}-\mathbf{L}_k^\mathrm{G}\right\|^2+H_\mathrm{U}^2\right) b_k N_0}\right)}, \tag{\ref{eq.ab}}\\
        \textrm{s.t.} \quad & \frac{D_k\left[\left(a_k^\mathrm{S}+a_k^\mathrm{U}\right)\rho_k+\left(1-a_k^\mathrm{S}-a_k^\mathrm{U}\right)\right]}{b_k\log_2\left(1+\frac{G_0 g_0 p_k}{\Theta^2 \left(\left\|\mathbf{L}^\mathrm{U}-\mathbf{L}_k^\mathrm{G}\right\|^2+H_\mathrm{U}^2\right) b_k N_0}\right)}\notag\\
        &\hspace{0.7in} \leq T-t_\mathrm{S}-t_\mathrm{SU}-t_k^\mathrm{U},\forall k\in\mathcal{K},\label{eq.ab.c1}\\
        & \left\|\mathbf{L}^\mathrm{U}-\mathbf{L}_k^\mathrm{G}\right\| \leq H_\mathrm{U}\tan\Theta,\forall k\in\mathcal{K},\label{eq.ab.c2}\\
        & H_\mathrm{U}^{\min}\leq H_\mathrm{U}\leq H_\mathrm{U}^{\max},\label{eq.ab.c3}\\
        & \Theta_{\min}\leq\Theta\leq\Theta_{\max}.\label{eq.ab.c4}
    \end{align}
\end{subequations}

We observe that the objective function of problem \eqref{eq.ab} and the left hand side of constraint \eqref{eq.ab.c1} are both increasing functions in $H_\mathrm{U}$ with given $\Theta$. Denote $H_\mathrm{U}^*$ as the optimal value of $H_\mathrm{U}$ in problem \eqref{eq.ab}, we can claim that
\begin{equation}\label{eq.oh}
    H_\mathrm{U}^*={\max}\left\{H_\mathrm{U}^{\min},\frac{L_{\max}}{\tan\Theta}\right\},
\end{equation}
where $L_{\max}={\max}_{k\in\mathcal{K}} \left\|\mathbf{L}^\mathrm{U}-\mathbf{L}_k^\mathrm{G}\right\|$. Based on \eqref{eq.oh}, we consider the following two cases.

\subsubsection{Case 1}
If $H_\mathrm{U}^*=H_\mathrm{U}^{\min}$, problem \eqref{eq.ab} is equivalent to
\begin{subequations}\label{eq.case1}
    \begin{align}
        \min_{\Theta} \quad & \Theta,\tag{\ref{eq.case1}}\\
        \textrm{s.t.} \quad & \frac{D_k\left[\left(a_k^\mathrm{S}+a_k^\mathrm{U}\right)\rho_k+\left(1-a_k^\mathrm{S}-a_k^\mathrm{U}\right)\right]}{b_k\log_2\left(1+\frac{G_0 g_0 p_k}{\Theta^2 \left[\left\|\mathbf{L}^\mathrm{U}-\mathbf{L}_k^\mathrm{G}\right\|^2+\left(H_\mathrm{U}^{\min}\right)^2\right] b_k N_0}\right)}\notag\\
        &\hspace{0.7in} \leq T-t_\mathrm{S}-t_\mathrm{SU}-t_k^\mathrm{U},\forall k\in\mathcal{K},\label{eq.case1.c1}\\
        & L_{\max} \leq H_\mathrm{U}^{\min}\tan\Theta,\label{eq.case1.c2}\\
        & \Theta_{\min}\leq\Theta\leq\Theta_{\max}.\label{eq.case1.c3}
    \end{align}
\end{subequations}
Obviously, the optimal solution of problem \eqref{eq.case1} is
\begin{equation}
    \Theta^*={\max}\left\{\Theta_{\min},\arctan \frac{L_{\max}}{H_\mathrm{U}^{\min}}\right\},
\end{equation}
which is the minimal value of $\Theta$ satisfying constraints \eqref{eq.case1.c2} and \eqref{eq.case1.c3}. Considering constraint \eqref{eq.case1.c1}, problem \eqref{eq.case1} is feasible if and only if
\begin{equation}
    \Theta^*\leq{\min}\left\{\Theta_{\max},\min_{k\in\mathcal{K}} \sqrt{\frac{G_0 g_0 p_k}{I_k b_k N_0 \left(2^{J_k}-1\right)}}\right\},
\end{equation}
where $I_k=\left\|\mathbf{L}^\mathrm{U}-\mathbf{L}_k^\mathrm{G}\right\|^2+\left(H_\mathrm{U}^{\min}\right)^2$ and
$J_k=\frac{D_k\left[\left(a_k^\mathrm{S}+a_k^\mathrm{U}\right)\rho_k+\left(1-a_k^\mathrm{S}-a_k^\mathrm{U}\right)\right]}{b_k(T-t_\mathrm{S}-t_\mathrm{SU}-t_k^\mathrm{U})}$.
Otherwise, problem \eqref{eq.case1} has no solution.

\subsubsection{Case 2}
If $H_\mathrm{U}^{\min}\geq\frac{L_{\max}}{\tan\Theta^*}$, then the optimal solution of case $1$ is the optimal solution of problem \eqref{eq.ab}. Otherwise, $H_\mathrm{U}^*=\frac{L_{\max}}{\tan\Theta}$. In this case, problem \eqref{eq.ab} is equivalent to
\begin{subequations}\label{eq.case2}
    \begin{align}
        \min_{\Theta} \quad & \sum_{k=1}^K \frac{p_k D_k\left[\left(a_k^\mathrm{S}+a_k^\mathrm{U}\right)\rho_k+\left(1-a_k^\mathrm{S}-a_k^\mathrm{U}\right)\right]}{b_k\log_2\left(1+\frac{G_0 g_0 p_k}{\Theta^2 \left[\left\|\mathbf{L}^\mathrm{U}-\mathbf{L}_k^\mathrm{G}\right\|^2+\left(\frac{L_{\max}}{\tan\Theta}\right)^2\right] b_k N_0}\right)},\tag{\ref{eq.case2}}\\
        \textrm{s.t.} \quad & \frac{D_k\left[\left(a_k^\mathrm{S}+a_k^\mathrm{U}\right)\rho_k+\left(1-a_k^\mathrm{S}-a_k^\mathrm{U}\right)\right]}{b_k\log_2\left(1+\frac{G_0 g_0 p_k}{\Theta^2 \left[\left\|\mathbf{L}^\mathrm{U}-\mathbf{L}_k^\mathrm{G}\right\|^2+\left(\frac{L_{\max}}{\tan\Theta}\right)^2\right] b_k N_0}\right)}\notag\\
        &\hspace{0.7in} \leq T-t_\mathrm{S}-t_\mathrm{SU}-t_k^\mathrm{U},\forall k\in\mathcal{K},\label{eq.case2.c1}\\
        & H_\mathrm{U}^{\min}\leq \frac{L_{\max}}{\tan\Theta}\leq H_\mathrm{U}^{\max},\label{eq.case2.c2}\\
        & \Theta_{\min}\leq\Theta\leq\Theta_{\max}.\label{eq.case2.c3}
    \end{align}
\end{subequations}
It is generally hard to obtain the optimal solution of problem \eqref{eq.case2} in closed form due to its complicated objective function. Hence, we conduct one-dimensional exhaustive search over $\left[\Theta_{\min},\Theta_{\max}\right]$ to obtain the optimal $\Theta^*$.

Comparing the optimal solution of the above two cases, the one with lower objective value is the optimal solution of problem \eqref{eq.ab}.

\subsection{Optimal Location Planning}\label{ssOLP}
With given satellite-UAV computation task allocation, semantic compression ratio, computation capacity, power, bandwidth allocation, altitude, and beamwidth, problem \eqref{eq.pf} can be simplified as
\begin{subequations}\label{eq.lp}
    \begin{align}
        \min_{\mathbf{L}^\mathrm{U}} \quad & \sum_{k=1}^K \frac{p_k D_k\left[\left(a_k^\mathrm{S}+a_k^\mathrm{U}\right)\rho_k+\left(1-a_k^\mathrm{S}-a_k^\mathrm{U}\right)\right]}{b_k\log_2\left(1+\frac{G_0 g_0 p_k}{\Theta^2 \left(\left\|\mathbf{L}^\mathrm{U}-\mathbf{L}_k^\mathrm{G}\right\|^2+H_\mathrm{U}^2\right) b_k N_0}\right)}, \tag{\ref{eq.lp}}\\
        \textrm{s.t.} \quad & \frac{D_k\left[\left(a_k^\mathrm{S}+a_k^\mathrm{U}\right)\rho_k+\left(1-a_k^\mathrm{S}-a_k^\mathrm{U}\right)\right]}{b_k\log_2\left(1+\frac{G_0 g_0 p_k}{\Theta^2 \left(\left\|\mathbf{L}^\mathrm{U}-\mathbf{L}_k^\mathrm{G}\right\|^2+H_\mathrm{U}^2\right) b_k N_0}\right)}\notag\\
        &\hspace{0.7in} \leq T-t_\mathrm{S}-t_\mathrm{SU}-t_k^\mathrm{U},\forall k\in\mathcal{K},\label{eq.lp.c1}\\
        & \left\|\mathbf{L}^\mathrm{U}-\mathbf{L}_k^\mathrm{G}\right\| \leq H_\mathrm{U}\tan\Theta,\forall k\in\mathcal{K}.\label{eq.lp.c2}
    \end{align}
\end{subequations}
Constraint \eqref{eq.lp.c1} is equivalent to
\begin{equation}\label{eq.lp.c1.e}
    \left\|\mathbf{L}^\mathrm{U}-\mathbf{L}_k^\mathrm{G}\right\| \leq \sqrt{\frac{G_0 g_0 p_k}{\Theta^2 b_k N_0 \left(2^{J_k}-1\right)}-H_\mathrm{U}^2},\forall k\in\mathcal{K}.
\end{equation}
Denote the right-hand side of \eqref{eq.lp.c1.e} by $Q_k$, then we can combine constraints \eqref{eq.lp.c1} and \eqref{eq.lp.c2} as
\begin{equation}\label{eq.lp.c1.2}
    \left\|\mathbf{L}^\mathrm{U}-\mathbf{L}_k^\mathrm{G}\right\| \leq {\min}\left\{H_\mathrm{U}\tan\Theta,Q_k\right\},\forall k\in\mathcal{K}.
\end{equation}
Since $\left\|\mathbf{L}^\mathrm{U}-\mathbf{L}_k^\mathrm{G}\right\|$ represents the horizontal distance between the UAV and GT $k$, the feasible region for GT $k$ is a circular area of radius ${\min}\left\{H_\mathrm{U}\tan\Theta,Q_k\right\}$ with center $\mathbf{L}_k^\mathrm{G}$. Denote the feasible region for GT $k$ by $\mathcal{R}_k$, we can express the feasible region of problem \eqref{eq.lp} as
\begin{equation}\label{eq.lp.fr}
    \mathcal{R}_\mathrm{U}=\bigcap_{k\in\mathcal{K}} \mathcal{R}_k.
\end{equation}

Then, we conduct two-dimensional exhaustive search over $\mathcal{R}_\mathrm{U}$ to obtain the optimal $\mathbf{L}^\mathrm{U}$ with the lowest objective value of problem \eqref{eq.lp}.

\subsection{Algorithm Analysis}
The overall SAGIN-enabled PSC network energy minimization algorithm is presented in Algorithm \ref{algo1}.

\begin{algorithm}[ht]
\caption{Energy Minimization Algorithm for SAGIN-enabled PSC Network}\label{algo1}
\begin{algorithmic}[1]
    \STATE Initialize $\left(\mathbf{L}^\mathrm{U}\right)^{(0)},H_\mathrm{U}^{(0)},\Theta^{(0)},\mathbf{b}^{(0)},\mathbf{f}^{(0)},\mathbf{p}^{(0)},\bm{\rho}^{(0)},$ $\left(\mathbf{a}^\mathrm{S}\right)^{(0)},\left(\mathbf{a}^\mathrm{U}\right)^{(0)}$. Set iteration index $i=1$.
    \REPEAT
        \STATE With given $\left(\mathbf{L}^\mathrm{U}\right)^{(i-1)},H_\mathrm{U}^{(i-1)},\Theta^{(i-1)},\mathbf{b}^{(i-1)},\mathbf{f}^{(i-1)},$ $\mathbf{p}^{(i-1)},\bm{\rho}^{(i-1)}$, solve the satellite-UAV computation task allocation subproblem and obtain the solution $\left(\mathbf{a}^\mathrm{S}\right)^{(i)},\left(\mathbf{a}^\mathrm{U}\right)^{(i)}$.
        \STATE With given $\left(\mathbf{L}^\mathrm{U}\right)^{(i-1)},H_\mathrm{U}^{(i-1)},\Theta^{(i-1)},\mathbf{b}^{(i-1)},\mathbf{f}^{(i-1)},$ $\mathbf{p}^{(i-1)},\left(\mathbf{a}^\mathrm{S}\right)^{(i)},\left(\mathbf{a}^\mathrm{U}\right)^{(i)}$, solve the semantic compression ratio optimization subproblem and obtain the solution $\bm{\rho}^{(i)}$.
        \STATE With given $\left(\mathbf{L}^\mathrm{U}\right)^{(i-1)},H_\mathrm{U}^{(i-1)},\Theta^{(i-1)},\mathbf{b}^{(i-1)},\mathbf{p}^{(i-1)},$ $\bm{\rho}^{(i)},\left(\mathbf{a}^\mathrm{S}\right)^{(i)},\left(\mathbf{a}^\mathrm{U}\right)^{(i)}$, solve the optimal computation capacity allocation subproblem and obtain the solution $\mathbf{f}^{(i)}$.
        \STATE With given $\left(\mathbf{L}^\mathrm{U}\right)^{(i-1)},H_\mathrm{U}^{(i-1)},\Theta^{(i-1)},\mathbf{f}^{(i)},\bm{\rho}^{(i)},$ $\left(\mathbf{a}^\mathrm{S}\right)^{(i)},\left(\mathbf{a}^\mathrm{U}\right)^{(i)}$, solve the optimal power and bandwidth allocation subproblem and obtain the solution $\mathbf{b}^{(i)},\mathbf{p}^{(i)}$.
        \STATE With given $\left(\mathbf{L}^\mathrm{U}\right)^{(i-1)},\mathbf{b}^{(i)},\mathbf{f}^{(i)},\mathbf{p}^{(i)},\bm{\rho}^{(i)},\left(\mathbf{a}^\mathrm{S}\right)^{(i)},$ $\left(\mathbf{a}^\mathrm{U}\right)^{(i)}$, solve the optimal altitude and beamwidth subproblem and obtain the solution $H_\mathrm{U}^{(i)},\Theta^{(i)}$.
        \STATE With given $H_\mathrm{U}^{(i)},\Theta^{(i)},\mathbf{b}^{(i)},\mathbf{f}^{(i)},\mathbf{p}^{(i)},\bm{\rho}^{(i)},\left(\mathbf{a}^\mathrm{S}\right)^{(i)},$ $\left(\mathbf{a}^\mathrm{U}\right)^{(i)}$, solve the optimal location planning subproblem and obtain the solution $\left(\mathbf{L}^\mathrm{U}\right)^{(i)}$.
        \STATE Set $i=i+1$.
    \UNTIL{the objective value of problem \eqref{eq.pf} converges.}
    \STATE \textbf{Output}: The optimized $\mathbf{L}^\mathrm{U},H_\mathrm{U},\Theta,\mathbf{b},\mathbf{f},\mathbf{p},\bm{\rho},\mathbf{a}^\mathrm{S},\mathbf{a}^\mathrm{U}$.
\end{algorithmic}
\end{algorithm}

\subsubsection{Convergence Analysis}
Denote the objective value of problem \eqref{eq.pf} at $i$-th iteration by $V_{\mathrm{obj}}^{(i)}$, and the objective value at $i$-th iteration after solving the first subproblem by $V_{\mathrm{s1}}^{(i)}$, etc. According to Algorithm \ref{algo1}, we have
\begin{multline}
    V_{\mathrm{obj}}^{(i-1)}\geq V_{\mathrm{s1}}^{(i)}\geq V_{\mathrm{s2}}^{(i)}\geq V_{\mathrm{s3}}^{(i)}\\
    \geq V_{\mathrm{s4}}^{(i)}\geq V_{\mathrm{s5}}^{(i)}\geq V_{\mathrm{s6}}^{(i)}=V_{\mathrm{obj}}^{(i)},
\end{multline}
which means the objective value of problem \eqref{eq.pf} is non-increasing along the iteration. Moreover, the physical meaning of the objective value is energy consumption, which is always positive. Since the objective value is non-increasing during the iteration and is lower-bounded by zero, the proposed iterative Algorithm \ref{algo1} must converge.

\subsubsection{Complexity Analysis}
According to Algorithm \ref{algo1}, the complexity of solving problem \eqref{eq.pf} lies in solving six subproblems at each iteration. For the satellite-UAV computation task allocation subproblem, the complexity is $\mathcal{O}\left(N_1 K\right)$, where $N_1$ is the number of iterations of using the dual method to solve problem \eqref{eq.sucta}.
For the semantic compression ratio optimization subproblem, the complexity lies in the segment selection problem and the subsequent convex problem. For the segment selection problem, the complexity is $\mathcal{O}\left(N_2 K D\right)$, where $N_2$ is the number of iterations of using the dual method to solve problem \eqref{eq.ss}. For the subsequent convex problem, the complexity is $\mathcal{O}\left(M_1^2 M_2\right)$ \cite{lobo1998applications}, where $M_1=K$ is the number of variables, and $M_2=2K$ is the number of constraints in problem \eqref{eq.scroass}. As a result, the total complexity of solving the semantic compression ratio optimization subproblem is $\mathcal{O}\left(N_2 K D+K^3\right)$.
For the optimal computation capacity allocation subproblem, the complexity is $\mathcal{O}\left(K\right)$.
For the optimal power and bandwidth allocation subproblem, the complexity is $\mathcal{O}\left(K^3\right)$.
For the optimal optimal altitude and beamwidth subproblem, the complexity is $\mathcal{O}\left(\left(\Theta_{\max}-\Theta_{\min}\right)/\eta\right)$, where $\eta$ is the step size of one-dimensional exhaustive search of problem \eqref{eq.case2}.
For the optimal location planning subproblem, the complexity is $\mathcal{O}\left(N_3\right)$, where $N_3$ is the number of steps of two-dimensional exhaustive search of problem \eqref{eq.lp}.
As a result, the total complexity of Algorithm \ref{algo1} is $\mathcal{O}(NN_1K+NN_2KD+NK^3+N\left(\Theta_{\max}-\Theta_{\min}\right)/\eta+NN_3)$, where $N$ is the number of outer iterations of Algorithm \ref{algo1}.

\section{Simulation Results and Analysis}\label{sra}
In the simulations, the GTs are uniformly distributed within a circular area of radius $300$ m. We assume that each GT requires the same amount of data.
For the PSC model, we adopt the same parameters as in \cite{ZHAO2024107055}. A summary of the main system parameters is provided in Table~\ref{tb1}.

\begin{table}[ht]
\centering
\caption{Main System Parameters}
\begin{tabular}{|c||c|}
    \toprule\hline
    \textbf{Parameter}  & \textbf{Value} \\
    \hline
    Number of GTs $K$ & $4$ \\ \hline
    Original data size $D_k$ & $64$ KB\\ \hline
    Satellite-UAV distance $d^\mathrm{SU}$ & $200$ km \\ \hline
    Satellite beam gain $\delta_\mathrm{S}$ & $25$ dB \\ \hline
    Satellite-UAV wavelength $\lambda^\mathrm{SU}$ & $10$ mm \\ \hline
    Satellite bandwidth $B_\mathrm{SU}$ & $1$ GHz \\ \hline
    Satellite transmit power $P_\mathrm{S}$ & $1$ W \\ \hline
    Power spectral density of AWGN $N_0$ & $-174$ dBm/Hz \\ \hline
    Reference gain $g_0$ & $1.42\times 10^{-4}$ \\ \hline
    Computation coefficient $\tau$ & $10^{-28}$ \\ \hline
    Satellite computation capacity $F_\mathrm{S}$ & $1$ GHz \\ \hline
    Maximum latency $T$ & $700$ ms \\ \hline
    Total transmit power of the UAV $P_\mathrm{U}$ & $1$ W \\ \hline
    Feasible altitude range of the UAV $\left[H_\mathrm{U}^{\min},H_\mathrm{U}^{\max}\right]$ & $\left[50,500\right]$ m \\ \hline
    Total bandwidth of the UAV $B_\mathrm{U}$ & $10$ MHz \\ \hline
    Total computation capacity of the UAV $F_\mathrm{U}$ & $0.5$ GHz \\ \hline
    Feasible range of half-beamwidth $\left[\Theta_{\min},\Theta_{\max}\right]$ & $\left[0,\pi/2\right]$ rad \\
    \hline\bottomrule
\end{tabular}
\label{tb1}
\end{table}

\begin{figure}[t]
\centering
\includegraphics[width=\linewidth]{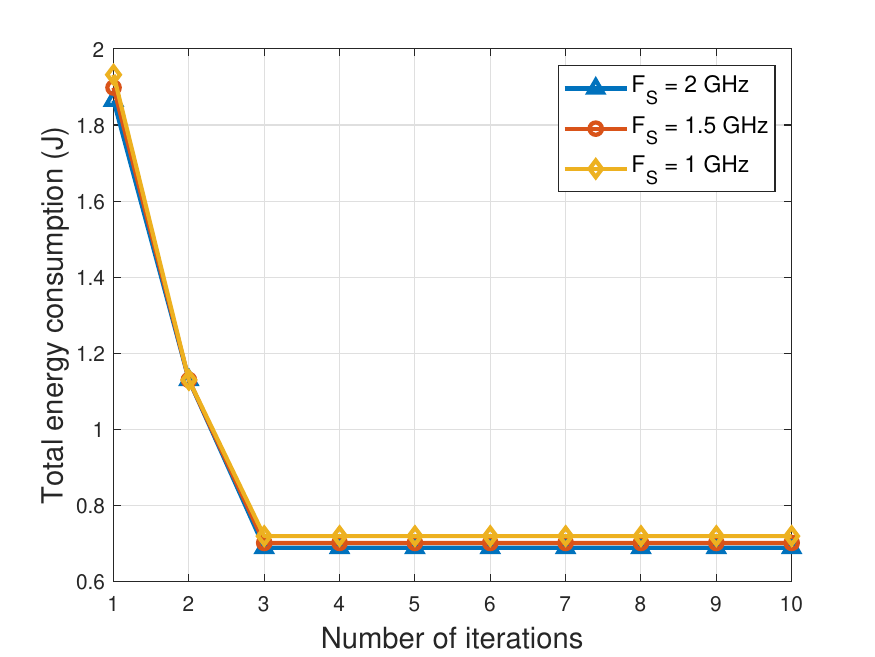}
\caption{Convergence behavior of the proposed algorithm under different satellite computation capacity.}
\label{fg.conv}
\end{figure}

Fig.~\ref{fg.conv} demonstrates the convergence behavior of the proposed algorithm under varying satellite computation capacities. The results indicate that the algorithm converges rapidly, requiring only three iterations to achieve stability, which underscores the effectiveness of our optimization algorithm. Initially, the energy consumption is high because communication and computation resources are equally allocated to each GT. However, after several iterations, the energy consumption significantly decreases, as the proposed algorithm effectively optimizes these system parameters.

To compare the results of the proposed algorithm, labeled as `SAGIN-PSC', we consider three alternative schemes: the `Non-semantic' scheme, which employs no semantic compression; the `Random comp. allocation' scheme, where computation tasks are randomly allocated between the satellite and the UAV; and the `Fix UAV location' scheme, which excludes optimization of the UAV's location.

\begin{figure}[t]
\centering
\includegraphics[width=\linewidth]{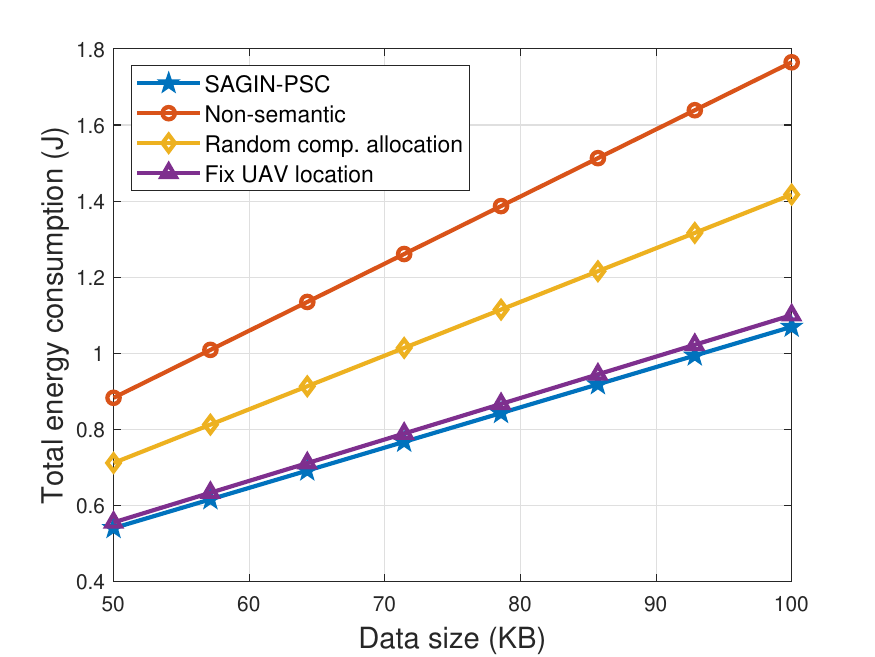}
\caption{Total energy consumption vs. data size.}
\label{fg.dk}
\end{figure}

Fig.~\ref{fg.dk} illustrates the relationship between total energy consumption and original data size. As expected, an increase in data size leads to a corresponding rise in energy consumption across all four examined schemes. Notably, the proposed `SAGIN-PSC' scheme consistently demonstrates the lowest total energy consumption. This efficiency is primarily attributed to the scheme's use of the PSC technique, which compresses the original data, therefore reducing the energy expended on communication. Despite the additional computational resources required for semantic compression, the energy savings from reduced communication outweigh the extra computational energy, especially when an optimal semantic compression ratio is applied. Additionally, the `SAGIN-PSC' scheme exhibits the gentlest slope in its energy consumption curve, underscoring its robustness in handling varying data sizes.

\begin{figure}[t]
\centering
\includegraphics[width=\linewidth]{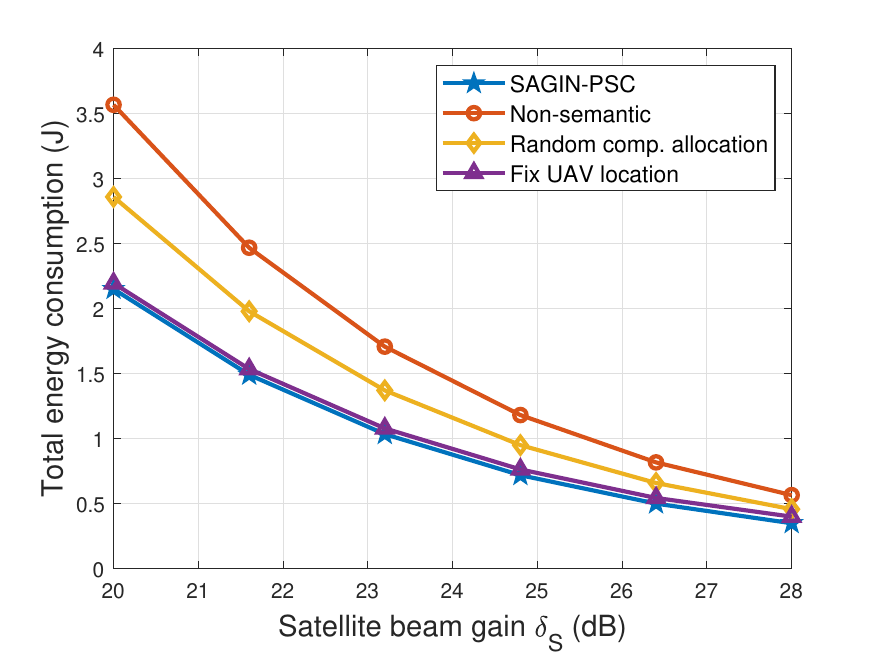}
\caption{Total energy consumption vs. satellite beam gain.}
\label{fg.ds}
\end{figure}

Fig.~\ref{fg.ds} demonstrates that the `SAGIN-PSC' scheme's benefits are particularly pronounced under conditions of low satellite beam gain. This occurs because a decrease in satellite beam gain diminishes the achievable rate between the satellite and the UAV, thereby increasing the energy required to transmit the same amount of data. Under these conditions, the `SAGIN-PSC' scheme compensates by allocating more computation resources at the satellite to mitigate the adverse effects of reduced beam gain.

\begin{figure}[t]
\centering
\includegraphics[width=\linewidth]{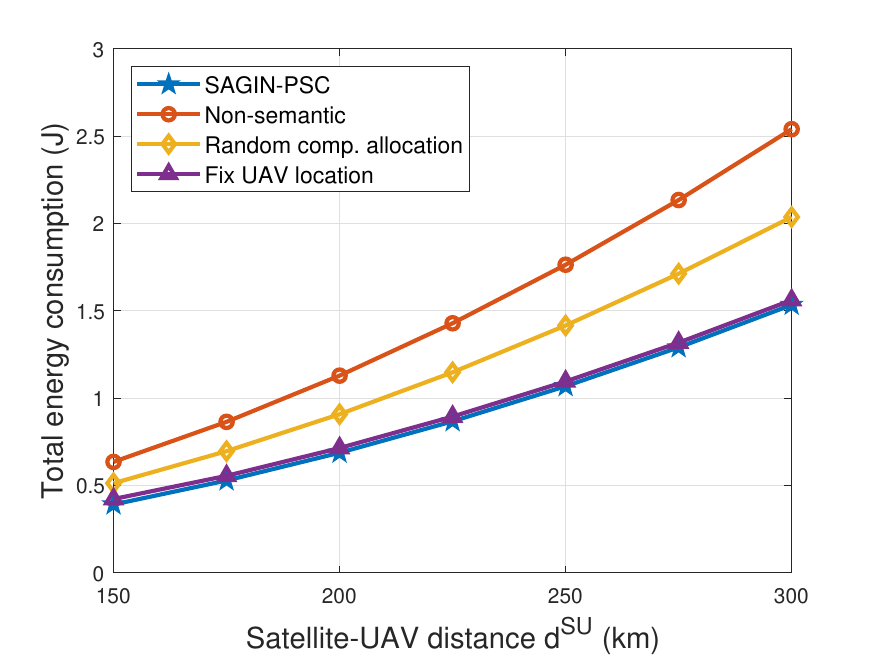}
\caption{Total energy consumption vs. satellite-UAV distance.}
\label{fg.dsu}
\end{figure}

Fig.~\ref{fg.dsu} depicts the relationship between total energy consumption and the distance between the satellite and UAV. As this distance increases, the channel gain between the satellite and UAV weakens, resulting in a lower achievable rate and consequently higher communication energy expenditure. This trend confirms that communication energy forms a significant component of the total energy consumption as distance increases. Interestingly, the `Fix UAV location' scheme exhibits energy consumption levels comparable to those of the SAGIN-PSC' scheme, suggesting that the UAV's location does not critically impact the system's overall energy efficiency. This observation also indicates that the energy used for communication between the UAV and GTs is relatively minor within the system's total energy consumption.

\begin{figure}[t]
\centering
\includegraphics[width=\linewidth]{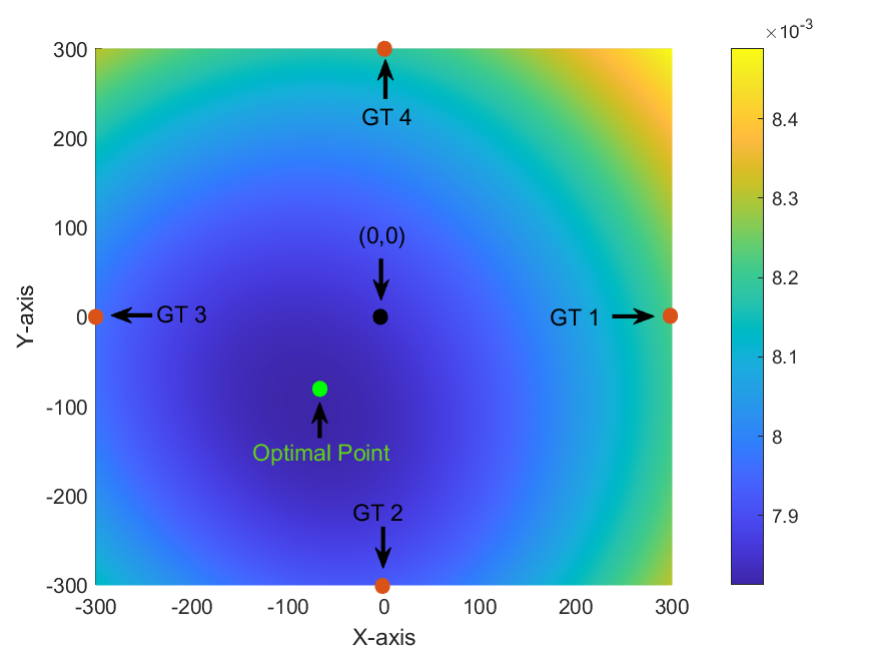}
\caption{The impact of UAV's location on UAV communication energy.}
\label{fg.loc}
\end{figure}

In Fig.~\ref{fg.loc}, the influence of UAV's location on its communication energy is explored. The axes represent the UAV's two-dimensional coordinates in meters, and the color gradient illustrates varying values of the objective function defined in Section \ref{ssOLP}, which is the communication energy consumed by the UAV. Fig.~\ref{fg.loc} is obtained when the $4$ GTs requiring different amount of data. The figure reveals that the optimal UAV location is not at the origin but slightly offset. Given that the UAV communication energy is quantified on the order of $10^{-3}$, it constitutes a minor fraction of the total system energy consumption. This minor impact supports the observation that the `Fix UAV location' scheme performs comparably to the `SAGIN-PSC' scheme.

\section{Conclusion}\label{c}
This paper has investigated the problem of energy efficiency in SAGIN-enabled PSC system. The model considers that a satellite transmits data to multiple GTs through a UAV acting as a relay. The satellite and the UAV can use PSC technique to compress the transmitted data, while the GTs can automatically recover missing information. The PSC is enabled by shared probability graphs among the transceivers, allowing for the conservation of communication resource at the expense of additional computation resource. The joint communication and computation problem is formulated as an optimization problem aiming to minimize the total communication and computation energy consumption of the network under latency, power, computation capacity, bandwidth, semantic compression ratio, and UAV location constraints. We proposed an iterative algorithm to solve this non-convex non-smooth problem, where the closed-form solutions for computation capacity allocation and UAV altitude are obtained at each iteration. Numerical results demonstrate the effectiveness of the proposed algorithm.

In future research, we plan to expand our scenario to include multiple satellites and UAVs. In addition, we plan to consider the dynamic elements of satellite motion and the trajectory of the UAV to improve the robustness and adaptability of our energy-efficient communication framework.

\bibliographystyle{IEEEtran}
\bibliography{main}

\begin{thebibliography}{10}
\providecommand{\url}[1]{#1}
\csname url@samestyle\endcsname
\providecommand{\newblock}{\relax}
\providecommand{\bibinfo}[2]{#2}
\providecommand{\BIBentrySTDinterwordspacing}{\spaceskip=0pt\relax}
\providecommand{\BIBentryALTinterwordstretchfactor}{4}
\providecommand{\BIBentryALTinterwordspacing}{\spaceskip=\fontdimen2\font plus
\BIBentryALTinterwordstretchfactor\fontdimen3\font minus \fontdimen4\font\relax}
\providecommand{\BIBforeignlanguage}[2]{{%
\expandafter\ifx\csname l@#1\endcsname\relax
\typeout{** WARNING: IEEEtran.bst: No hyphenation pattern has been}%
\typeout{** loaded for the language `#1'. Using the pattern for}%
\typeout{** the default language instead.}%
\else
\language=\csname l@#1\endcsname
\fi
#2}}
\providecommand{\BIBdecl}{\relax}
\BIBdecl

\bibitem{10080878}
R.~Samy, H.-C. Yang, T.~Rakia, and M.-S. Alouini, ``Space-air-ground {FSO} networks for high-throughput satellite communications,'' \emph{IEEE Commun. Mag.}, vol.~61, no.~3, pp. 82--87, Mar. 2023.

\bibitem{8368236}
J.~Liu, Y.~Shi, Z.~M. Fadlullah, and N.~Kato, ``Space-air-ground integrated network: A survey,'' \emph{IEEE Commun. Surv. Tutorials}, vol.~20, no.~4, pp. 2714--2741, May 2018.

\bibitem{9177315}
J.~Ye, S.~Dang, B.~Shihada, and M.-S. Alouini, ``Space-air-ground integrated networks: Outage performance analysis,'' \emph{IEEE Trans. Wireless Commun.}, vol.~19, no.~12, pp. 7897--7912, Dec. 2020.

\bibitem{10445211}
G.~Zheng, Q.~Ni, K.~Navaie, and H.~Pervaiz, ``Semantic communication in satellite-borne edge cloud network for computation offloading,'' \emph{IEEE J. Sel. Areas Commun.}, Feb. 2024.

\bibitem{9210567}
O.~Kodheli, E.~Lagunas, N.~Maturo, S.~K. Sharma, B.~Shankar, J.~F.~M. Montoya, J.~C.~M. Duncan, D.~Spano, S.~Chatzinotas, S.~Kisseleff, J.~Querol, L.~Lei, T.~X. Vu, and G.~Goussetis, ``Satellite communications in the new space era: A survey and future challenges,'' \emph{IEEE Commun. Surv. Tutorials}, vol.~23, no.~1, pp. 70--109, Jan. 2021.

\bibitem{9261433}
Y.~Zhu, W.~Bai, M.~Sheng, J.~Li, D.~Zhou, and Z.~Han, ``Joint {UAV} access and {GEO} satellite backhaul in {IoRT} networks: Performance analysis and optimization,'' \emph{IEEE Internet Things J.}, vol.~8, no.~9, pp. 7126--7139, May 2021.

\bibitem{8434289}
M.~Li, Y.~Hong, C.~Zeng, Y.~Song, and X.~Zhang, ``Investigation on the {UAV}-to-satellite optical communication systems,'' \emph{IEEE J. Sel. Areas Commun.}, vol.~36, no.~9, pp. 2128--2138, Sep. 2018.

\bibitem{9955992}
G.~Cui, P.~Duan, L.~Xu, and W.~Wang, ``Latency optimization for hybrid {GEO–LEO} satellite-assisted {IoT} networks,'' \emph{IEEE Internet Things J.}, vol.~10, no.~7, pp. 6286--6297, Apr. 2023.

\bibitem{9848831}
R.~Deng, B.~Di, H.~Zhang, H.~V. Poor, and L.~Song, ``Holographic {MIMO} for {LEO} satellite communications aided by reconfigurable holographic surfaces,'' \emph{IEEE J. Sel. Areas Commun.}, vol.~40, no.~10, pp. 3071--3085, Oct. 2022.

\bibitem{9393372}
A.~U. Chaudhry and H.~Yanikomeroglu, ``Laser intersatellite links in a {Starlink} constellation: A classification and analysis,'' \emph{IEEE Veh. Technol. Mag.}, vol.~16, no.~2, pp. 48--56, Jun. 2021.

\bibitem{8533634}
M.~Mozaffari, A.~Taleb Zadeh~Kasgari, W.~Saad, M.~Bennis, and M.~Debbah, ``Beyond {5G} with {UAVs}: Foundations of a {3D} wireless cellular network,'' \emph{IEEE Trans. Wireless Commun.}, vol.~18, no.~1, pp. 357--372, Jan. 2019.

\bibitem{8586877}
Q.~Song, F.-C. Zheng, Y.~Zeng, and J.~Zhang, ``Joint beamforming and power allocation for {UAV}-enabled full-duplex relay,'' \emph{IEEE Trans. Veh. Technol.}, vol.~68, no.~2, pp. 1657--1671, Feb. 2019.

\bibitem{8941314}
Z.~Yang, W.~Xu, and M.~Shikh-Bahaei, ``Energy efficient {UAV} communication with energy harvesting,'' \emph{IEEE Trans. Veh. Technol.}, vol.~69, no.~2, pp. 1913--1927, Dec. 2020.

\bibitem{8457275}
M.-N. Nguyen, L.~D. Nguyen, T.~Q. Duong, and H.~D. Tuan, ``Real-time optimal resource allocation for embedded {UAV} communication systems,'' \emph{IEEE Wireless Commun. Lett.}, vol.~8, no.~1, pp. 225--228, Feb. 2019.

\bibitem{8764580}
Z.~Yang, C.~Pan, K.~Wang, and M.~Shikh-Bahaei, ``Energy efficient resource allocation in {UAV}-enabled mobile edge computing networks,'' \emph{IEEE Trans. Wireless Commun.}, vol.~18, no.~9, pp. 4576--4589, Jul. 2019.

\bibitem{9955525}
D.~Gündüz, Z.~Qin, I.~E. Aguerri, H.~S. Dhillon, Z.~Yang, A.~Yener, K.~K. Wong, and C.-B. Chae, ``Beyond transmitting bits: Context, semantics, and task-oriented communications,'' \emph{IEEE J. Sel. Areas Commun.}, vol.~41, no.~1, pp. 5--41, Jan. 2023.

\bibitem{10550151}
Z.~Zhao, Z.~Yang, C.~Huang, L.~Wei, Q.~Yang, C.~Zhong, W.~Xu, and Z.~Zhang, ``A joint communication and computation design for distributed {RISs} assisted probabilistic semantic communication in {IIoT},'' \emph{IEEE Internet Things J.}, Jun. 2024.

\bibitem{9679803}
X.~Luo, H.-H. Chen, and Q.~Guo, ``Semantic communications: Overview, open issues, and future research directions,'' \emph{IEEE Wireless Commun.}, vol.~29, no.~1, pp. 210--219, Feb. 2022.

\bibitem{9955312}
W.~Yang, H.~Du, Z.~Q. Liew, W.~Y.~B. Lim, Z.~Xiong, D.~Niyato, X.~Chi, X.~Shen, and C.~Miao, ``Semantic communications for future internet: Fundamentals, applications, and challenges,'' \emph{IEEE Commun. Surv. Tutorials}, vol.~25, no.~1, pp. 213--250, Feb. 2023.

\bibitem{9530497}
G.~Shi, Y.~Xiao, Y.~Li, and X.~Xie, ``From semantic communication to semantic-aware networking: Model, architecture, and open problems,'' \emph{IEEE Commun. Mag.}, vol.~59, no.~8, pp. 44--50, Aug. 2021.

\bibitem{zhou2024feature}
K.~Zhou, G.~Zhang, Y.~Cai, Q.~Hu, G.~Yu, and A.~L. Swindlehurst, ``Feature allocation for semantic communication with space-time importance awareness,'' \emph{arXiv preprint arXiv:2401.14614}, Feb. 2024.

\bibitem{10024766}
W.~Xu, Z.~Yang, D.~W.~K. Ng, M.~Levorato, Y.~C. Eldar, and M.~Debbah, ``Edge learning for {B5G} networks with distributed signal processing: Semantic communication, edge computing, and wireless sensing,'' \emph{IEEE J. Sel. Topics Signal Process.}, vol.~17, no.~1, pp. 9--39, Jan. 2023.

\bibitem{10233741}
Z.~Zhao, Z.~Yang, Y.~Hu, L.~Lin, and Z.~Zhang, ``Semantic information extraction for text data with probability graph,'' in \emph{Proc. 2023 IEEE/CIC Int. Conf. Commun. China (ICCC Workshops)}, Aug. 2023.

\bibitem{9398576}
H.~Xie, Z.~Qin, G.~Y. Li, and B.-H. Juang, ``Deep learning enabled semantic communication systems,'' \emph{IEEE Trans. Signal Process.}, vol.~69, pp. 2663--2675, Apr. 2021.

\bibitem{zhao2024prob}
Z.~Zhao, Z.~Yang, Y.~Hu, Q.~Yang, W.~Xu, and Z.~Zhang, ``Probabilistic semantic communication over wireless networks with rate splitting,'' \emph{arXiv preprint arXiv:2403.00434}, Mar. 2024.

\bibitem{10183794}
Z.~Qin, F.~Gao, B.~Lin, X.~Tao, G.~Liu, and C.~Pan, ``A generalized semantic communication system: From sources to channels,'' \emph{IEEE Wireless Commun.}, vol.~30, no.~3, pp. 18--26, Jun. 2023.

\bibitem{9832831}
Y.~Wang, M.~Chen, T.~Luo, W.~Saad, D.~Niyato, H.~V. Poor, and S.~Cui, ``Performance optimization for semantic communications: An attention-based reinforcement learning approach,'' \emph{IEEE J. Sel. Areas Commun.}, vol.~40, no.~9, pp. 2598--2613, Sep. 2022.

\bibitem{9953316}
T.~Han, Q.~Yang, Z.~Shi, S.~He, and Z.~Zhang, ``Semantic-preserved communication system for highly efficient speech transmission,'' \emph{IEEE J. Sel. Areas Commun.}, vol.~41, no.~1, pp. 245--259, Jan. 2023.

\bibitem{9953076}
D.~Huang, F.~Gao, X.~Tao, Q.~Du, and J.~Lu, ``Toward semantic communications: Deep learning-based image semantic coding,'' \emph{IEEE J. Sel. Areas Commun.}, vol.~41, no.~1, pp. 55--71, Jan. 2023.

\bibitem{e26050394}
Z.~Zhao, Z.~Yang, M.~Chen, Z.~Zhang, and H.~V. Poor, ``A joint communication and computation design for probabilistic semantic communications,'' \emph{Entropy}, vol.~26, no.~5, Apr. 2024.

\bibitem{9763856}
L.~Yan, Z.~Qin, R.~Zhang, Y.~Li, and G.~Y. Li, ``Resource allocation for text semantic communications,'' \emph{IEEE Wireless Commun. Lett.}, vol.~11, no.~7, pp. 1394--1398, Jul. 2022.

\bibitem{10032275}
Z.~Yang, M.~Chen, Z.~Zhang, and C.~Huang, ``Energy efficient semantic communication over wireless networks with rate splitting,'' \emph{IEEE J. Sel. Areas Commun.}, vol.~41, no.~5, pp. 1484--1495, Jan. 2023.

\bibitem{10440193}
C.~Huang, G.~Chen, P.~Xiao, Y.~Xiao, Z.~Han, and J.~A. Chambers, ``Joint offloading and resource allocation for hybrid cloud and edge computing in {SAGINs}: A decision assisted hybrid action space deep reinforcement learning approach,'' \emph{IEEE J. Sel. Areas Commun.}, Feb. 2024.

\bibitem{10118916}
C.~Liu, C.~Guo, S.~Wang, Y.~Li, and D.~Hu, ``Task-oriented semantic communication based on semantic triplets,'' in \emph{Proc. 2023 IEEE Wireless Commun. Netw. Conf. (WCNC)}, Mar. 2023.

\bibitem{9039685}
J.~Li, A.~Sun, J.~Han, and C.~Li, ``A survey on deep learning for named entity recognition,'' \emph{IEEE Trans. Knowl. Data Eng.}, vol.~34, no.~1, pp. 50--70, Jan. 2022.

\bibitem{9446853}
Y.~Hu, H.~Shen, W.~Liu, F.~Min, X.~Qiao, and K.~Jin, ``A graph convolutional network with multiple dependency representations for relation extraction,'' \emph{IEEE Access}, vol.~9, pp. 81\,575--81\,587, Jun. 2021.

\bibitem{10226302}
W.~Wang, F.~Liu, W.~Liao, and L.~Xiao, ``Cross-modal graph knowledge representation and distillation learning for land cover classification,'' \emph{IEEE Trans. Geosci. Remote Sens.}, vol.~61, pp. 1--18, Aug. 2023.

\bibitem{10214643}
W.~Zhao and X.~Wu, ``Boosting entity-aware image captioning with multi-modal knowledge graph,'' \emph{IEEE Trans. Multimedia}, vol.~26, pp. 2659--2670, Mar. 2024.

\bibitem{9961954}
X.~Zhu, Z.~Li, X.~Wang, X.~Jiang, P.~Sun, X.~Wang, Y.~Xiao, and N.~J. Yuan, ``Multi-modal knowledge graph construction and application: A survey,'' \emph{IEEE Trans. Knowledge Data Eng.}, vol.~36, no.~2, pp. 715--735, Feb. 2024.

\bibitem{8788525}
X.~Zhang, X.~Sun, C.~Xie, and B.~Lun, ``From vision to content: Construction of domain-specific multi-modal knowledge graph,'' \emph{IEEE Access}, vol.~7, pp. 108\,278--108\,294, Aug. 2019.

\bibitem{10333452}
Z.~Zhao, Z.~Yang, Q.-V. Pham, Q.~Yang, and Z.~Zhang, ``Semantic communication with probability graph: A joint communication and computation design,'' in \emph{Proc. 2023 IEEE 98th Veh. Technol. Conf. (VTC2023-Fall)}, Oct. 2023.

\bibitem{ZHAO2024107055}
Z.~Zhao, Z.~Yang, X.~Gan, Q.-V. Pham, C.~Huang, W.~Xu, and Z.~Zhang, ``A joint communication and computation design for semantic wireless communication with probability graph,'' \emph{Journal of the Franklin Institute}, p. 107055, Jul. 2024.

\bibitem{balanis2016antenna}
C.~A. Balanis, \emph{Antenna theory: analysis and design}.\hskip 1em plus 0.5em minus 0.4em\relax New York, NY, USA: John wiley \& sons, 2016.

\bibitem{bertsekas2009convex}
D.~Bertsekas, \emph{Convex optimization theory}.\hskip 1em plus 0.5em minus 0.4em\relax Athena Scientific, Belmont, 2009.

\bibitem{lobo1998applications}
M.~S. Lobo, L.~Vandenberghe, S.~Boyd, and H.~Lebret, ``Applications of second-order cone programming,'' \emph{Linear Algebra Appl.}, vol. 284, no. 1-3, pp. 193--228, Nov. 1998.

\end{thebibliography}

\end{document}